\documentclass[twocolumn]{aastex631}
\usepackage{amsmath}

\newcommand{\mpch}{\>h^{-1}{\rm {Mpc}}}

\newcommand{\bmv}{\boldsymbol{v}}

\begin{document}

\title{DarkAI: Reconstructing the Density, Velocity and Tidal Fields of Dark Matter from a DESI-like Bright Galaxy Sample}

\correspondingauthor{Feng Shi}
\email{fshi@xidian.edu.cn}

\author[0000-0002-9968-2894]{Feng Shi}
\affiliation{School of Aerospace Science and Technology, Xidian University, Xi'an 710126, China}
\affiliation{Shaanxi Key Laboratory of Space Extreme Detection, Xidian University, Xi'an 710126, China}

\author{Zitong Wang}
\affiliation{School of Aerospace Science and Technology, Xidian University, Xi'an 710126, China}
\affiliation{Shaanxi Key Laboratory of Space Extreme Detection, Xidian University, Xi'an 710126, China}

\author[0000-0003-3997-4606]{Xiaohu Yang}
\affiliation{Tsung-Dao Lee Institute  \& School of Physics and Astronomy, Shanghai Jiao Tong University, Shanghai 200240, China}
% \affiliation{Key Laboratory for Particle Physics, Astrophysics and Cosmology (MoE), and Shanghai Key Laboratory for Particle\\ Physics and Cosmology, Shanghai Jiao Tong University, Shanghai {\rm 200240}, China}
\affiliation{State Key Laboratory of Dark Matter 
Physics, Key Laboratory for Particle Astrophysics and Cosmology (MOE), \& 
Shanghai Key Laboratory for Particle Physics and Cosmology, Shanghai Jiao Tong 
University, Shanghai 200240, China}

\author{Yizhou Gu}
\affiliation{Tsung-Dao Lee Institute  \& School of Physics and Astronomy, Shanghai Jiao Tong University, Shanghai 200240, China}
% \affiliation{Key Laboratory for Particle Physics, Astrophysics and Cosmology (MoE), and Shanghai Key Laboratory for Particle\\ Physics and Cosmology, Shanghai Jiao Tong University, Shanghai {\rm 200240}, China}
\affiliation{State Key Laboratory of Dark Matter 
Physics, Key Laboratory for Particle Astrophysics and Cosmology (MOE), \& 
Shanghai Key Laboratory for Particle Physics and Cosmology, Shanghai Jiao Tong 
University, Shanghai 200240, China}

\author{Chengliang Wei}
\affiliation{Purple Mountain Observatory, Nanjing 210008, China}

\author[0000-0002-1318-4828]{Ming Li}
\affiliation{National Astronomical Observatories, Chinese Academy of Sciences, Beijing 100101, China}

\author[0000-0002-8010-6715]{Jiaxin Han}
\affiliation{Department of Astronomy, Shanghai Jiao Tong University, Shanghai 200240, China}
% \affiliation{Key Laboratory for Particle Physics, Astrophysics and Cosmology (MoE), and Shanghai Key Laboratory for Particle\\ Physics and Cosmology, Shanghai Jiao Tong University, Shanghai {\rm 200240}, China}
\affiliation{State Key Laboratory of Dark Matter 
Physics, Key Laboratory for Particle Astrophysics and Cosmology (MOE), \& 
Shanghai Key Laboratory for Particle Physics and Cosmology, Shanghai Jiao Tong 
University, Shanghai 200240, China}

\author{Zhejie Ding}
\affiliation{University of Chinese Academy of Sciences, Nanjing 211135, China}

\author[0000-0002-4911-6990]{Huiyuan Wang}
\affiliation{Department of Astronomy, University of Science and Technology of China, Hefei, Anhui 230026, China}

\author[0000-0003-1967-4091]{Youcai Zhang}
\affiliation{Shanghai Astronomical Observatory, Nandan Road 80, Shanghai 200030, China}

\author{Wensheng Hong}
\affiliation{Department of Astronomy, Shanghai Jiao Tong University, Shanghai 200240, China}

\author[0000-0003-3203-3299]{Yirong Wang}
\affiliation{Department of Astronomy, Shanghai Jiao Tong University, Shanghai 200240, China}

\author[0000-0003-3964-0438]{Xiao-dong Li}
\affiliation{School of Physics and Astronomy, Sun Yat-Sen University, Zhuhai 519082, China}
\affiliation{CSST Science Center for the Guangdong–Hong Kong–Macau Greater Bay Area, SYSU, Zhuhai 519082, China}

\newcommand{\zd}[1]{\textcolor{red}{[\textbf{ZD}: #1}]}

%% Note that the \and command from previous versions of AASTeX is now
%% depreciated in this version as it is no longer necessary. AASTeX 
%% automatically takes care of all commas and "and"s between authors names.

%% AASTeX 6.31 has the new \collaboration and \nocollaboration commands to
%% provide the collaboration status of a group of authors. These commands 
%% can be used either before or after the list of corresponding authors. The
%% argument for \collaboration is the collaboration identifier. Authors are
%% encouraged to surround collaboration identifiers with ()s. The 
%% \nocollaboration command takes no argument and exists to indicate that
%% the nearby authors are not part of surrounding collaborations.

%% Mark off the abstract in the ``abstract'' environment. 
\begin{abstract}
Reconstructing the mass density, velocity, and tidal (MTV) fields of dark matter from galaxy surveys is essential for advancing our understanding of the LSS of the Universe. In this work, we present a machine learning-based framework using a UNet convolutional neural network to reconstruct the MTV fields from mock samples of the DESI bright galaxy survey within the redshift range $0.1 < z < 0.4$. Our approach accounts for realistic observational effects, including geometric selection, flux-limited data, and redshift space distortion (RSD)  effects, thereby improving the fidelity of the reconstructed fields.  Testing on mock galaxy catalogs generated from the Jiutian N-body simulation, our method achieves significant accuracy level.  The reconstructed density field exhibits strong consistency with the true field, effectively eliminating most RSD effects and achieving a cross-correlation power spectrum coefficient greater than 0.985 on scales with $k < 0.1 \, h \, \mathrm{Mpc}^{-1}$. The velocity field reconstruction accurately captures large-scale coherent flows and small-scale turbulent features, exhibiting slopes of grid-to-grid relationships close to unity and scatter below $\sim$100 $\mathrm{km} \, \mathrm{s}^{-1}$. Additionally, the tidal field is reconstructed without bias, successfully recovering the features of the large-scale cosmic web, including clusters, filaments, sheets, and voids.  Our results confirm that the proposed framework effectively captures the large-scale distribution and dynamics of dark matter while addressing key systematic challenges. These advancements provide a reliable and robust tool for analyzing current and future galaxy surveys, paving the way for new insights into cosmic structure formation and evolution.

\end{abstract}

%% Keywords should appear after the \end{abstract} command. 
%% The AAS Journals now uses Unified Astronomy Thesaurus concepts:
%% https://astrothesaurus.org
%% You will be asked to selected these concepts during the submission process
%% but this old "keyword" functionality is maintained in case authors want
%% to include these concepts in their preprints.

%% From the front matter, we move on to the body of the paper.
%% Sections are demarcated by \section and \subsection, respectively.
%% Observe the use of the LaTeX \label
%% command after the \subsection to give a symbolic KEY to the
%% subsection for cross-referencing in a \ref command.
%% You can use LaTeX's \ref and \label commands to keep track of
%% cross-references to sections, equations, tables, and figures.
%% That way, if you change the order of any elements, LaTeX will
%% automatically renumber them.
%%
%% We recommend that authors also use the natbib \citep
%% and \citet commands to identify citations.  The citations are
%% tied to the reference list via symbolic KEYs. The KEY corresponds
%% to the KEY in the \bibitem in the reference list below. 

\section{Introduction} \label{sec:intro}

Understanding the distribution and dynamics of matter in the Universe provides critical insights into gravitational interactions and cosmic flows.  This knowledge is essential for exploring the processes that drive the formation and evolution of cosmic structures and galaxies. To achieve this, galaxy redshift surveys, such as the 2-degree-field Galaxy Redshift Survey \citep{2001MNRAS.328.1039C} and the Sloan Digital Sky Survey \citep{2000AJ....120.1579Y}, offer an opportunity to probe the large-scale mass distribution of the Universe. In the coming decade, next-generation galaxy surveys, including the Dark Energy Spectroscopic Instrument (DESI) \citep{2016arXiv161100036D,2016arXiv161100037D}, the Vera C. Rubin Observatory's Legacy Survey of Space and Time \citep{2019ApJ...873..111I}, EUCLID \citep{2011arXiv1110.3193L}, the Nancy Grace Roman Space Telescope \citep{2019arXiv190205569A}, and the Chinese Space Station Telescope \citep[e.g.,][]{2011SSPMA..41.1441Z,2019ApJ...883..203G,2025SCPMA..6849511S}, are poised to map an unprecedented volume of the Universe with different redshift measurement accuracies. These surveys will enable transformative advances in our understanding of cosmic structure and evolution, emphasizing the importance of developing robust and efficient methods for reconstructing the three-dimensional mass density, velocity, and tidal (MTV) fields of dark matter from the vast galaxy survey datasets.

Reconstructing the MTV field from observational data is essential for addressing key scientific questions. Constrained simulations, such as those in the local Universe \citep[e.g.,][]{2016ApJ...831..164W,2017ApJ...841...55T}, require accurate initial conditions derived from the density field to reproduce observed structures. Local non-Gaussianity in the cosmic web can be quantified through the MTV field, offering insights into primordial fluctuations \citep[][]{2010CQGra..27l4011D,2018arXiv181013423S, 2019PhRvD.100h3508M}. Filament spin studies, such as the alignment of the galaxy angular momentum with filaments \citep[e.g.,][]{2013ApJ...775L..42T,2021NatAs...5..839W}, rely on the reconstructed tidal field for precise measurements. Filament lensing studies, mapping dark matter in filaments \citep[e.g.,][]{2024NatAs...8..377H}, are similarly dependent on accurate reconstructions. Moreover, correcting supernova distance measurements for mass density and peculiar velocities \citep{2006PhRvD..73j3002C, 2006PhRvD..73l3526H, 2011ApJ...741...67D, 2024arXiv241006053H} and mapping baryon distributions using the velocity field \citep[e.g.,][]{2013MNRAS.432.1600D, 2019ApJ...887..265Y} underscore the broad utility of MTV reconstructions.

Reconstructing the cosmic density and velocity fields has previously been
conducted from several redshift surveys \citep{1995MNRAS.272..885F,1995ApJ...449..446Z,1999AJ....118.1146S,2002MNRAS.333..739M,2004MNRAS.352..939E,2009MNRAS.394..398W,2013ApJ...772...63W}. Early efforts \citep{1995MNRAS.272..885F, 1995ApJ...449..446Z} relied on smoothing and normalizing galaxy data to approximate large-scale density fields, with methods such as Wiener reconstruction minimizing errors through linear combinations of galaxy density values. Other approaches \citep{2009MNRAS.394..398W,2012MNRAS.420.1809W,2013ApJ...772...63W} leveraged dark matter halos or galaxy groups, assuming a linear bias between halo and mass density fields. However, these methods are quite complicated. The assumption of a scale-independent bias parameter lacks robust physical justification and is often motivated by simplicity, reducing its reliability in capturing complex relationships \citep[e.g.,][]{2007JCAP...10..007C,2018A&A...613A..15S}. Moreover, linear theory assumptions, which posit a simple proportionality between gravity and velocity fields, fail in the nonlinear clustering regime, due to phenomena like shell crossing, where the one-to-one correspondence between density and velocity fields breaks down. Additionally, redshift space distortions \citep[RSDs][]{1977ApJ...212L...3S,1983ApJ...267..465D,1987MNRAS.227....1K,1992ApJ...385L...5H}, influenced by the Kaiser effect on large scales \citep{1987MNRAS.227....1K} and the Finger-of-God effect on small scales \citep{1972MNRAS.156P...1J,1978IAUS...79...31T}, intertwine complex velocity and density field distortions, further challenging accurate reconstructions. Methods based on linear theory or the Zel'dovich approximation are particularly limited in high-density regions \citep[e.g.,][]{2016ApJ...833..241S, 2018ApJ...861..137S}, emphasizing the need for innovative strategies to fully leverage redshift survey data in mapping the Universe's LSS (LSS).

%  A UNet-based neural network was used to reconstruct peculiar velocity fields from redshift-space distributions of dark matter halos, achieving high accuracy at multiple scales

Several studies have employed machine learning (ML) techniques to reconstruct dark matter density or velocity fields, demonstrating significant potential for advancing the field \citep{2021ApJ...913....2W,2023arXiv230104586W,2023MNRAS.523.6272C,2023JCAP...06..062Q, 2024SCPMA..6719513W, 2024arXiv241111280X}.  However, these approaches face notable limitations. First, these approaches often operate within idealized simulation environments, with limitations such as fixed box volumes, periodic boundary conditions, and the exclusion of observational systematics like geometric selection and flux-limited effects. Second, reconstruction based on the halo distributions compared to the galaxies requires additional bias modeling to bridge the gap between the galaxies and dark matter. Addressing these limitations is essential for extending the ML-based reconstruction to real-world cosmological observations.

In this paper, we aim to develop a robust ML-based framework to reconstruct the MTV fields from a DESI-like bright galaxy survey (BGS). In \citet[][hereafter Wang24]{2024SCPMA..6719513W}, the first paper in this series, we presented a deep learning technique for reconstructing the mass density field from the redshift-space distribution of dark matter halos. The reconstruction was clearly proved to be able to correct for RSDs and was unaffected by the different cosmologies between the training and test samples. It was also tested that the application of a UNet-reconstructed density field can accurately infer the velocity and tidal field, providing better results compared to a traditional approach based on a linear bias model. Here, we improve upon the reconstruction method of Wang24 by addressing realistic observation challenges, including accounting for geometric selection, varying redshift distributions, and flux-limited observational biases. By incorporating systematic effects into the ML model, we aim to enhance the fidelity of the reconstructed fields and expand their applicability to real survey data. This study concentrates solely on reconstruction using the distribution of galaxies. We will address reconstruction involving groups or halos derived from observational data in a future probe.

The structure of this paper is as follows. Section~\ref{sec:data} provides an overview of the data and mock galaxy catalogs used for training and testing. Section~\ref{sec:meth} describes the ML-based reconstruction framework, including network architecture and training protocols. Section~\ref{sec:res} evaluates the performance of reconstruction on mock DESI-like catalogs, addressing systematic effects and observational biases. Finally, Section~\ref{sec:sum} presents a summary and prospects for future work.

\section{data} \label{sec:data}

To support the DESI galaxy survey, we generate mock galaxy samples along with their corresponding mass density and velocity fields of dark matter within a DESI-like light cone. These mock data sets are essential for studying the mapping between galaxies and the underlying LSS of the Universe, as well as for testing and validating cosmological analysis pipelines. The mock samples are constructed through a multistep process that includes N-body simulations, halo and subhalo identifications, galaxy population modeling, and the incorporation of survey selection effects. This section outlines the methodology utilized to generate these mock samples, based on the procedures described in \citet{2024MNRAS.529.4015G}.

The foundation of our mock galaxy samples is the Jiutian N-body simulation \citep{2025arXiv250321368H}, which was run at the High-Performance Computing Center at Kunshan using L-GADGET, a memory-optimized version of the GADGET2 code \citep{2005MNRAS.364.1105S}. The Jiutian simulation describes the distribution of $6144^3$ dark matter particles in a periodic box of $1000 \mpch$, with a particle mass of $3.723 \times 10^8 h^{-1} M_\odot$. The cosmological parameters were set to be consistent with the Planck2018 results \citep{2020A&A...641A...6P}: $\Omega_m = 0.3111$, $\Omega_{\Lambda} = 0.6889$, $\Omega_b = 0.049$, $\sigma_8 = 0.8102$, and $n_s = 0.9665$.

Dark matter halos are identified using the friends-of-friends algorithm \citep{1985ApJ...292..371D} with a linking length of 0.2 times the mean interparticle separation. Subhalos and their evolutionary histories are further processed using the Hierarchical Bound-Tracing (HBT+) algorithm \citep{2012MNRAS.427.2437H, 2018MNRAS.474..604H}, which operates in the time domain, to track the evolution of each halo throughout the simulation. The minimum number of particles in a subhalo is set to 20, corresponding to a minimum halo mass of approximately $7.5 \times 10^9 h^{-1} M_\odot$. The HBT+ code provides high-quality subhalo catalogs and a robust merger tree by following the most bound particle, even if a subhalo is no longer resolved. 

We populate the dark matter halos in our simulation with mock galaxies using the subhalo abundance matching method. This technique establishes a connection between the properties of galaxies and their corresponding subhalos, by assuming a monotonic relationship between these properties \citep{2004ApJ...609...35K, 2004MNRAS.353..189V, 2006ApJ...647..201C, 2012ApJ...752...41Y,2018ARA&A..56..435W}. The positions and velocities of the galaxies are assigned by directly inheriting the positions and velocities of their host subhalos from the N-body simulations. The luminosities are simulated by matching the $z$-band cumulative galaxy luminosity functions measured from the DESI One-percent survey \citep{2024AJ....168...58D} with the cumulative subhalo mass functions derived from the peak mass of all the subhalos. To reflect the luminosity-subhalo mass relation more realistically, a scatter in the $z$-band luminosity, $\sigma_\mathrm{log}(L_z)=0.15$ dex \citep[e.g.][]{2008ApJ...676..248Y}, is added to each galaxy.
\begin{figure*}
    \centering
    \includegraphics[trim=0cm 0cm 0cm 0cm, clip=True,width=1.8\columnwidth]{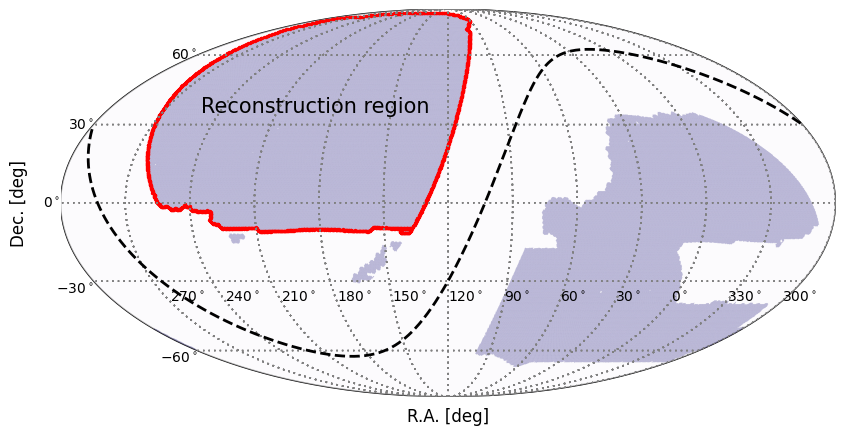}
    \caption{The footprint of DESI Legacy Imaging Survey DR9. The red contour indicates the region selected for reconstruction. The dashed black line represents the Galactic plane, excluding galaxies with $|b| < 25$ in Galactic coordinates.}
    \label{fig:skycov}
\end{figure*}
\begin{figure}
    \centering
    \includegraphics[trim=0cm 0cm 0cm 0cm, clip=True,width=1.\columnwidth]{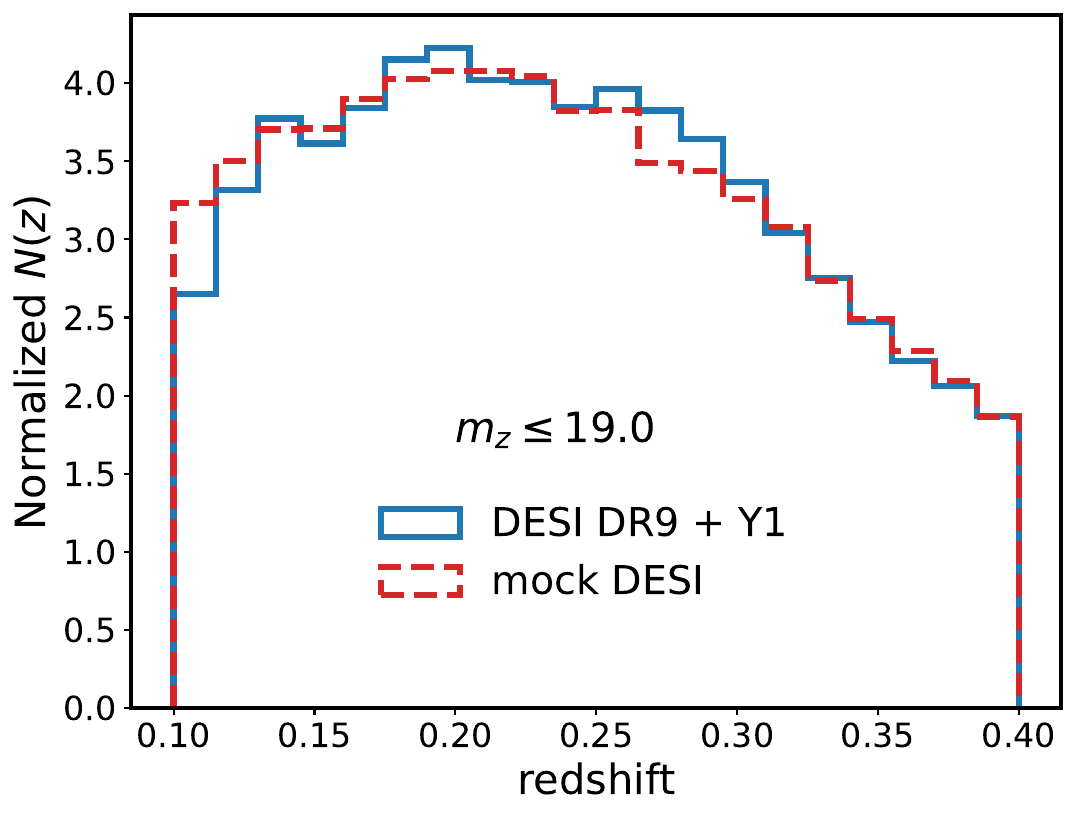}
    \caption{Normalized redshift distribution. The red dashed line shows the distribution for our mock DESI sample. The blue solid line shows the DESI galaxy sample, constructed by crossmatching photometric data from DESI LS DR9 with Year 1 spectroscopic observations.}
    \label{fig:nz}
\end{figure}
\begin{figure*}
    \centering
    \includegraphics[trim=0cm 0cm 0cm 0cm, clip=True,width=2.1\columnwidth]{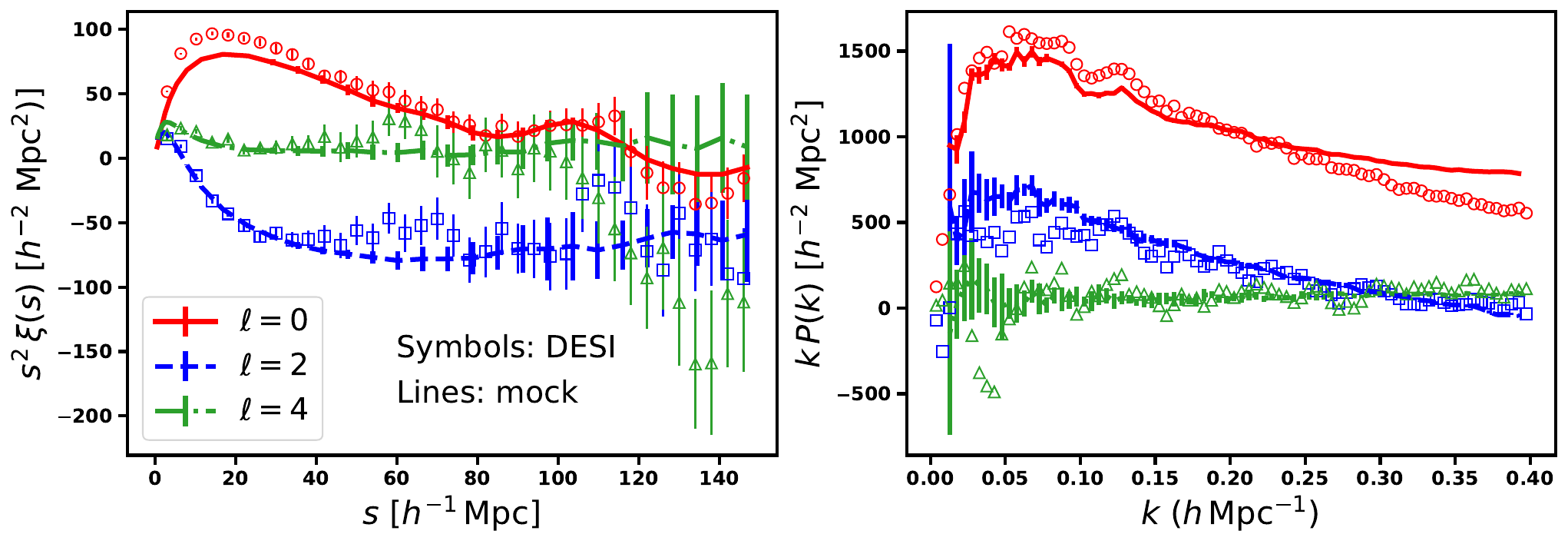}
    \caption{Comparison of galaxy clustering between mock samples and DESI BGS measurements. The left panel displays the multipole moments of the 2PCF, while the right panel shows the corresponding power spectra. The lines represent results from the mock samples, while the symbols denote the DESI BGS measurements.}
    \label{fig:xipk_comp}
\end{figure*}

We construct a mock DESI survey and its corresponding dark matter particle distribution using the following steps:

\begin{enumerate}
    \item We place a virtual observer at the center of the populated simulation box and stack replicas of the box to generate a past light cone. This process involves selecting snapshots from the N-body simulations that intersect with the light-cone volume at different redshifts.
    
    \item We define an equatorial coordinate system and remove all mock galaxies or dark matter particles located outside the DESI LS DR9 footprint, as shown in Figure~\ref{fig:skycov}. We also apply a foreground mask to exclude regions containing globular clusters, planetary nebulae, nearby large galaxies, and Gaia stars with $G < 16$. Additionally, we remove the region near the Galactic plane, $|b| < 25$, to avoid areas of high stellar density.
    
    \item For galaxies, we assign each a redshift, $z_{\rm obs}$, including the effects of RSD, calculated as:
    \begin{equation}
        z_{\rm obs} = z_{\rm cos} + z_{\rm pec} = z_{\rm cos} + \frac{v_{\rm pec}}{c}(1+z_{\rm cos}),
    \end{equation}
    where $z_{\rm cos}$ is the cosmological redshift from Hubble expansion, $z_{\rm pec}$ is the Doppler contribution due to the galaxy's line-of-sight peculiar velocity, $v_{\rm pec}$, and $c$ is the speed of light. For dark matter particles, we retain them in real space.
    
    \item For galaxies, we assign a $z$-band apparent magnitude, $m_z$, based on their distance and luminosity, incorporating a $K$-correction,  described as:
    \begin{equation}
        K^{0.5}(z) = 0.73z^2 - 0.54z - 0.33.
    \end{equation}
\end{enumerate}

In this study, we focus on the BGS sample, with a $z$-band magnitude limit cut $m_z<19.0$ and within the redshift range of $0.1<z<0.4$, which is consistent with DESI DR1 BGS \citep{2024arXiv240403000D}. Figure~\ref{fig:nz} shows the normalized redshift distributions for our mock DESI sample (red dashed line). For comparison, we also include the redshift distribution derived from the DESI galaxy sample (blue solid line), which is generated by \citet{2024ApJ...971..119W}. This sample is based on photometric imaging from the DESI Legacy Imaging Surveys (LS) DR9, with initial redshift estimates provided by the photometric redshifts for the LS catalog \citep{2021MNRAS.501.3309Z}. For galaxies with available spectra from DESI Year 1, redshifts are updated using {\tt fastspecfit} measurements, adopting the recommended “best” redshift in cases of multiple observations. To match the selection criteria of BGS, we further restrict the sample to galaxies with $m_z < 19.0$ and $0.1<z<0.4$. The redshift distribution of our mock sample generally follows the observed trend. In addition, DESI will cover two sky regions: a larger region in the Northern Galactic Cap (NGC) and a smaller region in the Southern Galactic Cap. To use a more contiguous region, we apply our reconstruction to the main region of the NGC, as shown in the red contour of Figure~\ref{fig:skycov}. The final sky coverage used in our reconstruction is about 9625.23 deg$^2$. In this study, we construct a set of 12 mock samples by randomly rotating and shifting the stacked simulation boxes, which are used for training and testing our network. 

% Since the DESI Y1 catalogue covers a relatively small fraction of the sky, a slight difference between the two distributions is expected due to cosmic variance. 
% \scr{are 10 mocks enough?}
%

To evaluate the consistency between our mock galaxy samples and the DESI observations, Figure~\ref{fig:xipk_comp} shows the comparison of the galaxy clustering. The left panels display the multipole moments of the two-point correlation function (2PCF), while the right panels show the corresponding power spectra. The lines represent the mock results, and the symbols denote the DESI results \citep{2024arXiv241112020D}, measured from the BGS  with $0.1 < z < 0.4$ and $M_r < -21.5$. The same selection criteria are applied to the mock sample to ensure consistency. Overall, both the 2PCF and power spectra exhibit reasonable agreement between the mock and DESI samples, indicating that the mock sample broadly reproduces the observed large-scale distributions. However, noticeable discrepancies remain, particularly in the monopole, primarily because we have not applied any clustering fitting in the mock sample relative to the observational data. We plan to incorporate such clustering fitting in future analyses, using the more robust DESI DR2 or DR3 data releases. Nevertheless, the current comparison demonstrates that our mock sample follows the DESI clustering distribution well and provides a reliable basis for testing the reconstruction framework and evaluating systematic effects.

\begin{figure*}
    \centering
    \includegraphics[trim=0cm 0cm 0cm 0cm, clip=True,width=2.1\columnwidth]{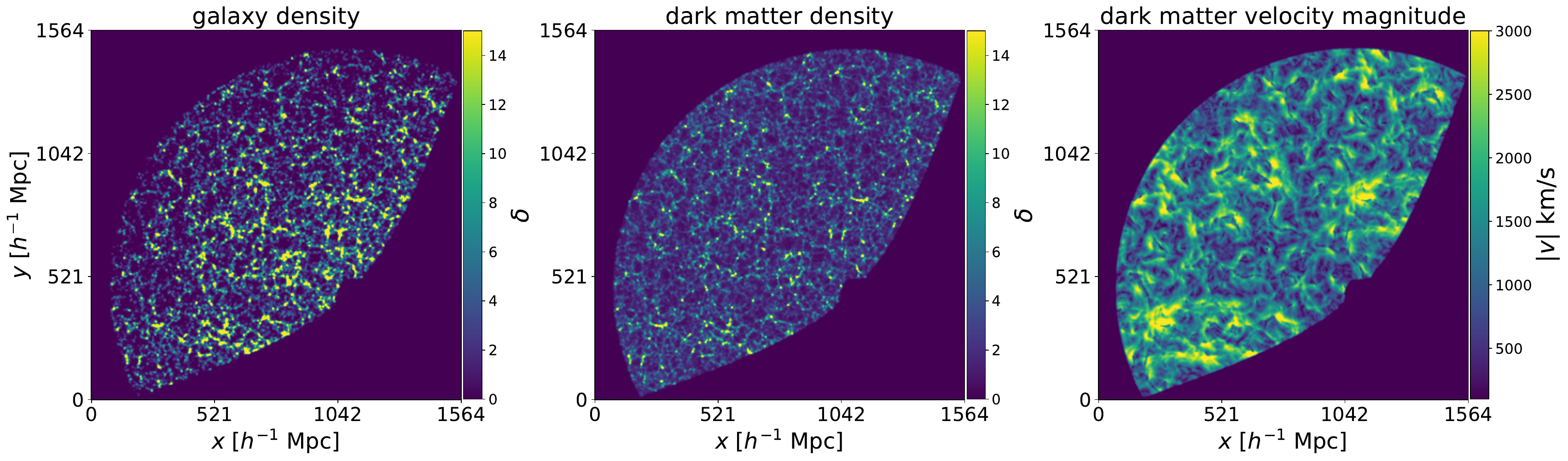}
    \caption{The projected field distributions in a slice of $1564 \times 1564 \times 9.16 \, h^{-1} \, \mathrm{Mpc}$. From left to right, the panels correspond to the galaxy overdensity, dark matter density, and velocity magnitude $|\boldsymbol{v}|$, respectively. }
    \label{fig:slices_gdv}
\end{figure*}
To construct the grid-based fields for dark matter and galaxies within the DESI-like survey volume, we proceed as follows. First, we embed the survey volume within a periodic cubic box with a side length of $1564 \, h^{-1} \, \mathrm{Mpc}$. To maximize the utilization of the box volume, we apply a constant angular rotation in the R.A. direction. The 3D positions $(x, y, z)$ of the objects are then calculated by
\begin{equation}
\begin{aligned}
x &= r \cos(\beta) \cos(\alpha - \alpha_\mathrm{rot}), \\
y &= r \cos(\beta) \sin(\alpha - \alpha_\mathrm{rot}), \\
z &= r \sin(\beta),
\end{aligned}
\end{equation}
where $r$ is the line-of-sight comoving distance, $\beta$ is the declination (Dec.), $\alpha$ is the R.A., and $\alpha_\mathrm{rot}=58^\circ$ is the applied rotation angle in this study.

We use the cloud-in-cell (CIC) scheme \citep{1981csup.book.....H} to compute the overdensity field, $\delta$, on a grid of $512^3$ grids for both dark matter and galaxy distributions. The velocity field, $\bmv$, is also estimated using the CIC method and normalized by the number of dark matter particles in each cell of the grid. To suppress the shot noise, we smooth both the $\delta$ and $\boldsymbol{v}$ fields using a TensorFlow convolutional kernel, with a kernel size of 3. Figure~\ref{fig:slices_gdv} shows the projected field distributions in a slice of $1564 \times 1564 \times 9.16 \, h^{-1} \, \mathrm{Mpc}$. From left to right, the panels correspond to the galaxy overdensity, dark matter density, and velocity magnitude $|\boldsymbol{v}|$, respectively. The correlated structures among the three fields are clearly visible.

In this study, we focus on evaluating the reconstruction performance using mock DESI samples, accounting for geometric selection, flux limits, and RSDs. However, another systematic effect, namely the fiber assignment (FA), can also introduce biases into the reconstruction. We include a brief discussion of this effect in Appendix \ref{sec:FA}. However, a comprehensive analysis of the FA effect and its correction is beyond the scope of this paper and will be addressed in future work.

\section{METHODOLOGY}\label{sec:meth}
In this section, we present our reconstruction method based on the UNet convolutional neural network. 

\subsection{Reconstruction via Maximum Likelihood Estimation}\label{sec:MLE}

Reconstructing the mass density and velocity fields from the galaxy density field constitutes a field-to-field mapping problem, which can be framed as a maximum likelihood estimation task \citep[e.g.][]{2021MNRAS.501.1499M}. The objective is to infer the underlying dark matter fields, given the observed galaxy density field. Assuming a probabilistic relationship between the two, the likelihood function is given by
\begin{equation}
\mathcal{L} = \prod_{k=1}^{N_p} P(\phi_{\mathrm{m},k} | \delta_{\mathrm{g},k}; \boldsymbol{\theta}),
\end{equation}
where $k$ indexes the grids, $N_p$ denotes the total number of grids, $\phi_{\mathrm{m}}$ represents the dark matter field (e.g., $\delta_\mathrm{m}$ or $\boldsymbol{v}_\mathrm{m}$), $\delta_{\mathrm{g}}$ is the galaxy density field, and $\boldsymbol{\theta}$ represents the model parameters. This formulation assumes that the grids are statistically independent, thereby simplifying the problem to a mapping between the corresponding grid points in the fields.

The estimation of $\boldsymbol{\theta}$ can be expressed in the following log-likelihood form:

% \scb{The estimation of $ \boldsymbol{\theta} $ simplifies to the following log-likelihood form,}

\begin{equation}\label{eq:loglike}
\hat{\boldsymbol{\theta}}_{\mathrm{ML}} = \underset{\theta}{\mathrm{arg\,min}} \sum_{k} \left[f(\delta_{\mathrm{g},k};\boldsymbol{\theta}) - \phi_{\mathrm{m},k} \right]^2,
\end{equation}
where $ f(\delta_{\mathrm{g},k};\boldsymbol{\theta}) $ represents the predicted field value assuming a Gaussian distribution in the data. This shows that maximizing the likelihood is equivalent to minimizing the mean squared error (MSE) between the true and predicted fields.

% making the log-likelihood proportional to the sum of squared differences between the predicted field $ f(\delta_{\mathrm{g},k};\boldsymbol{\theta}) $ and the true field $ \phi_{\mathrm{m},k} $. Consequently, maximizing the likelihood is equivalent to minimizing the Mean Squared Error (MSE), providing a statistically grounded and computationally efficient framework for parameter estimation.

% \scb{where $f(\delta_{\mathrm{g},k};\boldsymbol{\theta})$ represents the predicted field value given a Gaussian distribution, modeled as a function of the observed galaxy density field $ \delta_{\mathrm{g},k} $ and the parameter set $ \boldsymbol{\theta} $. This shows that maximizing the likelihood is equivalent to minimizing the Mean Squared Error (MSE) between the true and predicted fields.}

However, solving this formulation analytically is impractical, due to the model's complexity. Traditional optimization methods face challenges in estimating $\boldsymbol{\theta}$ due to the high dimensionality of the parameter space. Alternatively, neural networks provide an efficient approach by maximizing likelihood through supervised learning, effectively encoding $\boldsymbol{\theta}$ in their weights and biases.

% %
\begin{figure*}
    % \vspace{-7ex}%
	\hspace*{0.5cm}\includegraphics[width=2.1\columnwidth]{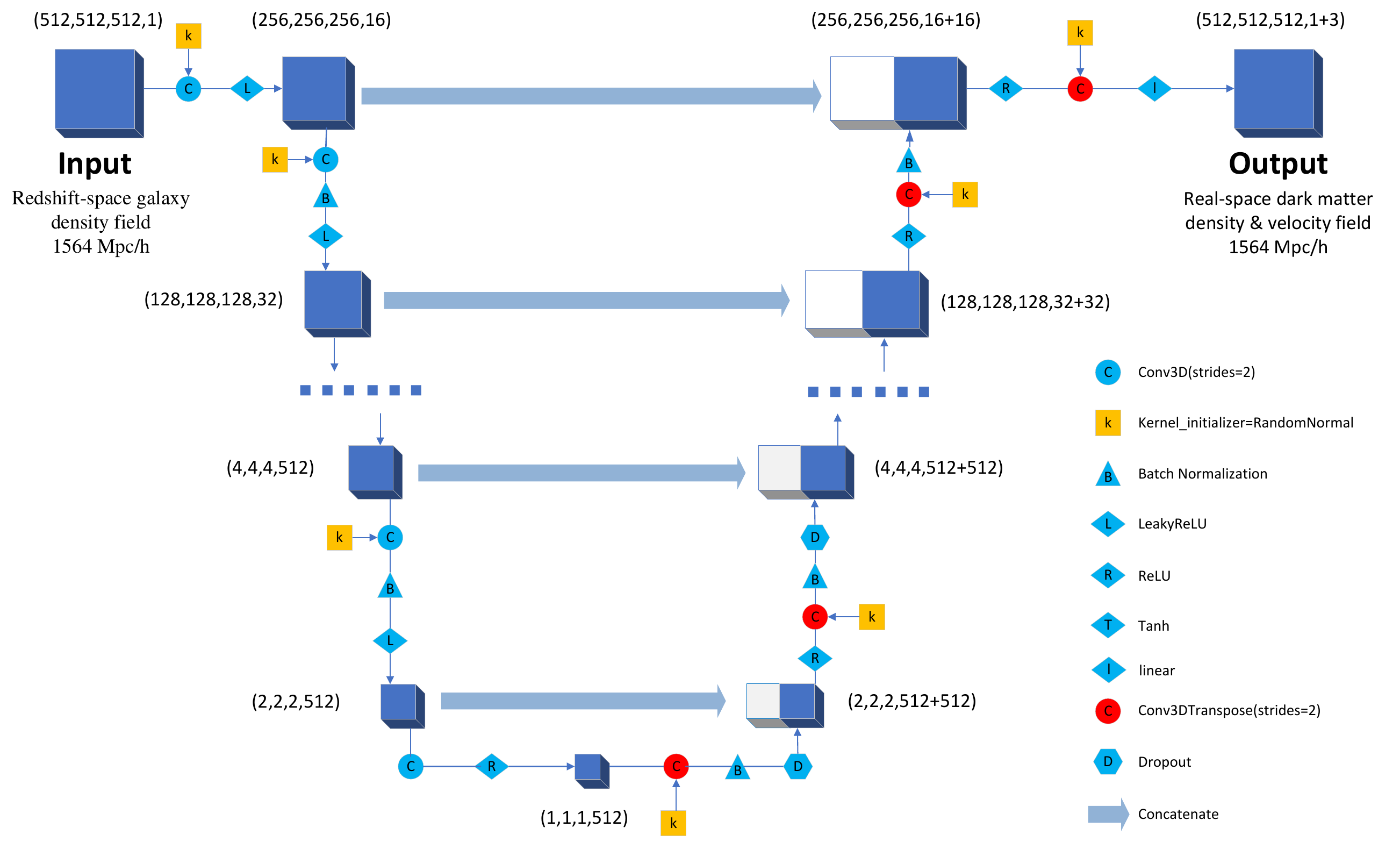}
    \caption{3D UNet architecture. Each cube represents a 3D space-with-feature map, with the grid size and the number of feature channels indicated. The input and output cubes, each containing $512^3$ voxels and spanning $1564 \, \mpch$, correspond to the redshift-space galaxy density field and the real-space dark matter density or velocity field, respectively. The encoder comprises 8 convolutional layers that progressively downsample the input cube to a bottleneck representation, while the decoder uses 8 transposed convolutional layers to upsample the bottleneck output to the target cube. The channel configuration (1+3) corresponds to the density field, $\delta_\mathrm{m}$, and the velocity components $(v_x, v_y, v_z)$, enabling simultaneous reconstruction of these quantities as an optional feature. The computational steps are represented by distinct symbols, as illustrated in the lower right corner. All convolutional layers utilize $4 \times 4 \times 4$ filters with a stride of $2$.}
    \label{fig:network}
\end{figure*}

\subsection{UNet Architecture}

We base our network architecture on the UNet style \citep{2015arXiv150504597R}, which is particularly well suited to reconstructing complex fields, due to its hierarchical convolutional layers and skip connections. These design features allow the model to capture multiscale spatial information effectively, which is critical for learning the mapping between the redshift-space galaxy density field and the target fields.

To implement this architecture, we adapt the generator design from the Pix2Pix framework \citep{2016arXiv161107004I}, tailoring it to our specific requirements. As illustrated in Figure~\ref{fig:network}, the network processes input and output cubes, each comprising $512^3$ grids. The input cube represents the redshift-space galaxy density field $\delta_\mathrm{g}$, while the output cube corresponds to either the real-space dark matter density field $\delta_\mathrm{m}$ or the velocity field $\boldsymbol{v}_\mathrm{m}$. The mapping is achieved through a symmetric encoder-decoder structure, where all convolutional layers utilize $4 \times 4 \times 4$ filters with a stride of $2$. The architecture includes the following key components:
\begin{itemize}
    \item \textbf{Encoder Path}: The encoder extracts features from the input field, $\delta_\mathrm{g}$, while progressively reducing the spatial resolution. Each block consists of a sequence: convolution, batch normalization, and LeakyReLU activation. Batch normalization is omitted in the first and last layers of the encoder, where the last layer uses ReLU activation instead of LeakyReLU.
    
    \item \textbf{Decoder Path}: The decoder reconstructs the target field incrementally, by increasing spatial resolution. Each block consists of a sequence: transposed convolution, batch normalization, dropout (with a rate of 50\%), and ReLU activation. Batch normalization is not applied in the first layer of the decoder, and the output layer employs a linear activation function.
    
    \item \textbf{Skip Connections}: Skip connections link the encoder layers to their corresponding decoder layers, enabling the preservation of spatial details during downsampling and upsampling.
    
    \item \textbf{Output Layer}: For the density field reconstruction, the output layer generates a single-channel output corresponding to $\delta_\mathrm{m}$. For the velocity field reconstruction, the output layer produces three channels, representing the velocity components $(v_x, v_y, v_z)$.
\end{itemize}

In this architecture, the fields $\delta_\mathrm{m}$ and $\boldsymbol{v}_\mathrm{m}$ are reconstructed separately, due to memory limitations, while they could be reconstructed simultaneously, by designing a multichannel output for the network hardware permitting. A detailed comparison between the separately trained and unified models is provided in Appendix~\ref{sec:compare_mods}.

\begin{figure*}
    \centering
    \includegraphics[trim=0cm 0cm 0cm 0cm, clip=True,width=2.\columnwidth]{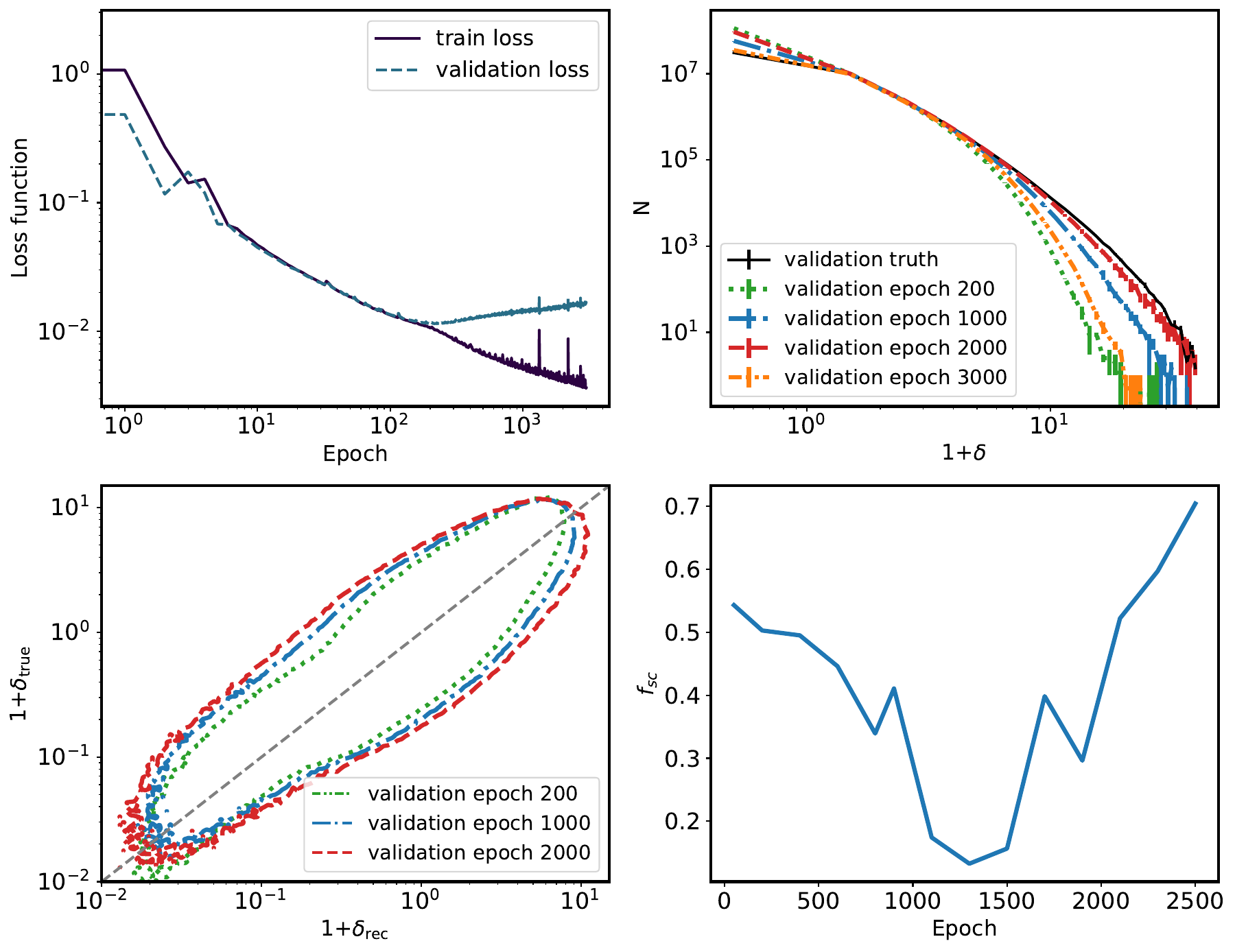}
    \caption{ Upper left: loss function for training $(\delta_{\mathrm{g}}, \delta_{\mathrm{m}})$ using 3000 epochs.  The dark solid and blue dashed lines represent the loss functions from the training and validation samples, respectively. Upper right: comparison of the density distributions for the validation samples. The black solid line represents the result of the target validation samples. Different colored lines represent the predicted fields at epochs 200, 1000, 2000, and 3000, repectively. Lower left: grid-to-grid density relation between the reconstructed density field $1+\delta_\mathrm{rec}$ and the true field $1+\delta_\mathrm{true}$. The contour in each case represents threshold encompassing 99\% of the grid cells. The dashed lines indicate the ideal one-to-one correspondence. The different colored curves correspond to different epochs, as indicated. Lower right: the evolution of the factor $f_{sc}$, defined as the product of the scatter variance and one minus the correlation scope between $\delta_\mathrm{rec}$ and $\delta_\mathrm{true}$, as a function of epoch. }
    \label{fig:train_loss}
\end{figure*}
\subsection{Training Process}\label{sec:train}

As described in Section~\ref{sec:data}, we generate paired data samples $ (\delta_{\mathrm{g}}, \delta_{\mathrm{m}}) $ and $ (\delta_{\mathrm{g}}, \boldsymbol{v}_{\mathrm{m}}) $ to train the networks. These samples represent the galaxy density field $ \delta_{\mathrm{g}} $ paired with the true density field $ \delta_{\mathrm{m}} $ or the velocity field $ \boldsymbol{v}_{\mathrm{m}} $. The objective is to learn the mapping between the galaxy and dark matter fields.

The training process then minimizes the MSE loss function, which corresponds to the log-likelihood function in Equation~(\ref{eq:loglike}).  For practical implementation, the loss function is expressed as 
\begin{equation}
\mathcal{L'} = \frac{1}{N_p} \sum_{k=1}^{N_p} \left(\phi^\mathrm{pred}_{\mathrm{m}, k} - \phi^\mathrm{true}_{\mathrm{m}, k}\right)^2,
\end{equation}
where $\phi^\mathrm{pred}_{\mathrm{m}} $ is the field being reconstructed and $ \phi^\mathrm{true}_{\mathrm{m}} $ denotes the true target field. Specifically, $ \phi_{\mathrm{m}} \equiv \delta_{\mathrm{m}} $ for density reconstruction and $ \phi_{\mathrm{m}} \equiv \boldsymbol{v}_{\mathrm{m}} $ for velocity reconstruction. 

To improve the training stability and optimize the network's ability to capture variations across a wide range of values, we rescale the input fields before training. The overdensity field is rescaled using a logarithmic transformation:  
\begin{equation}
    y = \log_{10}(1+\delta + a),
\end{equation}
where the parameter $a = 0.02$ ensures numerical stability, by preventing singularities for low-density regions ($\delta \approx -1$). Moreover, the $a$ value is optimized in our analysis to balance the reconstruction, enhancing the accuracy in high-density regions while preserving reliable performance in low-density areas. The velocity field is rescaled linearly as 
\begin{equation}
    y = \frac{\boldsymbol{v}_\mathrm{m}}{b},
\end{equation}
where the parameter $b = 3000 \, \mathrm{km} \, \mathrm{s}^{-1}$ normalizes the velocity values to a comparable scale, avoiding excessively large gradients during training. These rescaling transformations enhance the network's convergence and improve its performance, by standardizing the input dynamic range.

The network is trained from scratch using the {\tt Adam} optimizer \citep{2017arXiv171105101L}, which incorporates the adaptive learning rate. Each epoch processes five training samples, followed by two validation samples, used exclusively for model evaluation during training.

When training neural networks, selecting an optimal stopping point is critical to balancing generalization and overfitting, particularly for fields with wide-ranging density distributions. As shown in the upper left panel of Figure~\ref{fig:train_loss}, the validation loss converges around epoch 200, indicating an appropriate point for terminating training. However, relying solely on loss metrics may overlook the model performance across varying density regions, especially in high-density areas, which represent a small fraction of the total volume.

To address this limitation, we compare the density distributions of the predicted and true fields (upper right panel of Figure~\ref{fig:train_loss}). From epoch 200 to 2000, the disparities in high-density regions with $(1+\delta) > 6$ progressively diminish, indicating significant improvements in recovering high-density structures even after loss convergence. However, as training progresses, predictions exhibit systematic underestimation at epoch 3000, revealing overfitting effects.

However, the grid-to-grid relationship (lower left panel of Figure~\ref{fig:train_loss}) supports the opposite trend. From epoch 200 to 2000, the scatter increases while maintaining minimal bias, indicating improved density distribution recovery at the cost of reduced correlation with the true field. This trade-off highlights the need for a quantitative criterion to identify the optimal training stage.

We define the factor $f_{sc}$ as the product of the perpendicular variance, represented by the smaller eigenvalue of the covariance matrix of $ (\vec{\delta}_\mathrm{rec}, \vec{\delta}_\mathrm{true}) $, and the absolute value of 1 minus the correlation slope, which is given by the slope of the eigenvector corresponding to the larger variance. As illustrated in the lower right panel of Figure~\ref{fig:train_loss}, this factor reaches its minimum value at approximately epoch 1300 and increases significantly after epoch 2000. Consequently, halting the training process around epoch 1300 achieves a balance between high-density recovery and correlation retention, ensuring the reconstructed field accurately represents cosmological structures.

These results underscore the necessity of carefully assessing the training stages to ensure the reliability of the model for cosmological inference. The methodology introduced here could be generalized to other datasets \citep[e.g.][]{2024PhRvD.109f3509S}, offering a pathway to robust ML applications in astrophysical studies.
\begin{figure*}
    \centering
    \includegraphics[trim=0cm 0cm 0cm 0cm, clip=True,width=2.1\columnwidth]{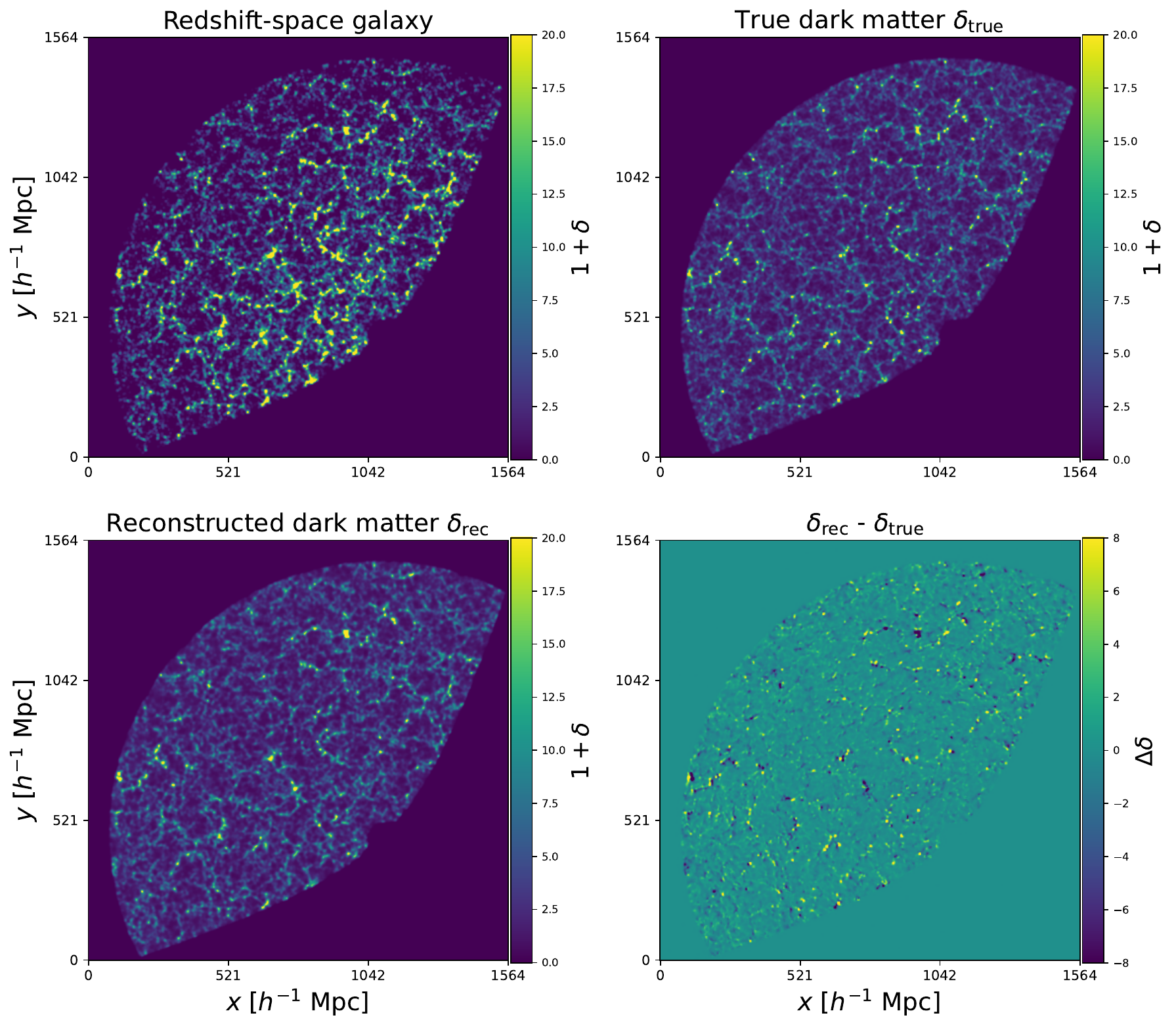}
    \caption{ Comparison of the projected density field distributions in a slice of $1564 \times 1564 \times 9.16 \, h^{-1} \, \mathrm{Mpc}$. The different panels correspond to the redshift-space galaxy distribution (input), the true dark matter density field $\delta_\mathrm{true}$ (target), the UNet-reconstructed density field $\delta_\mathrm{rec}$ (output), and the difference between $\delta_\mathrm{rec}$ and $\delta_\mathrm{true}$, as labeled at the top of each panel.}
    \label{fig:slice_den}
\end{figure*}
\begin{figure*}
    \centering
    \includegraphics[trim=0cm 0cm 0cm 0cm, clip=True,width=2.\columnwidth]{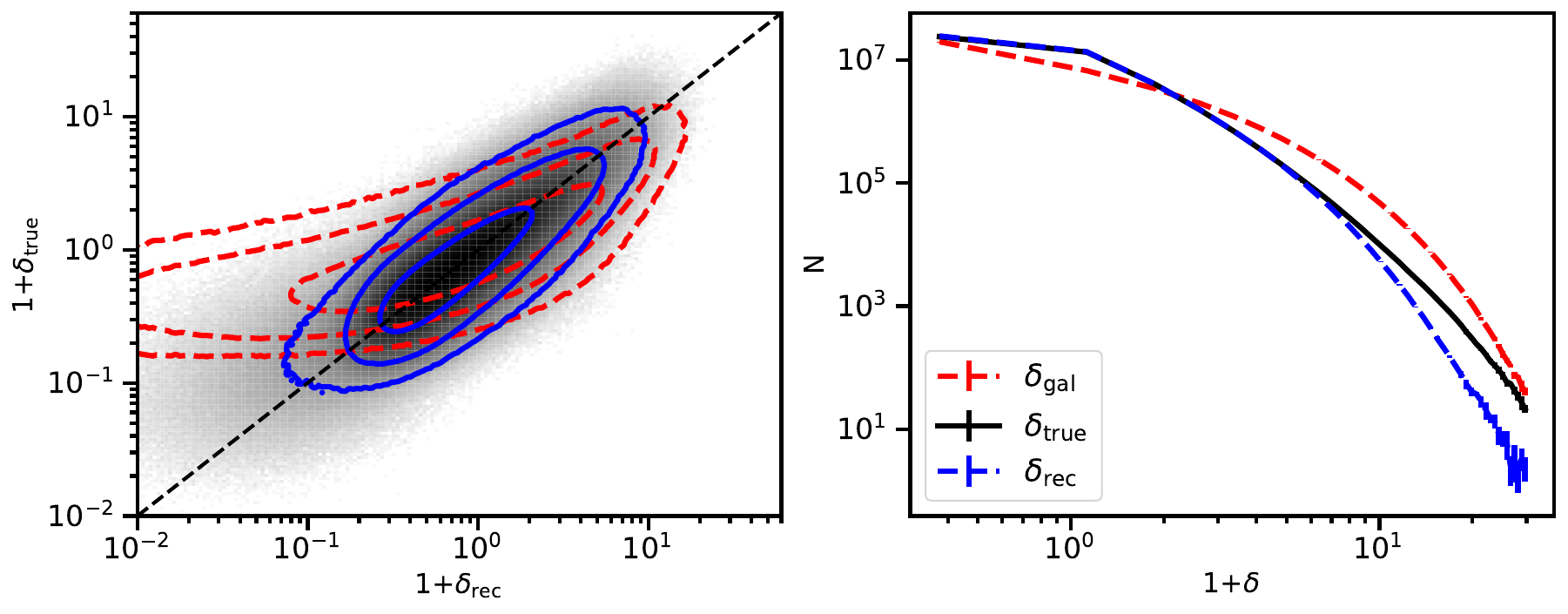}
    \caption{ Left: grid-to-grid density relationship between the reconstructed density field $1+\delta_\mathrm{rec}$ and the true field $1+\delta_\mathrm{true}$, represented by the blue solid line. For comparison, the result for the galaxy density field is also plotted with the red dashed line. The three contours in each case represent thresholds encompassing 67\%, 95\%, and 99\% of the grid cells. The dashed lines indicate the ideal one-to-one correspondence. Right: density distributions for galaxies (red dashed line), the reconstructed field (blue solid line), and the true field (black solid line). The error bars indicate the $\pm 1\sigma$ variance among the five test samples. }
    \label{fig:pix_com}
\end{figure*}
\begin{figure*}
    \centering
    \includegraphics[trim=0cm 0cm 0cm 0cm, clip=True,width=2.\columnwidth]{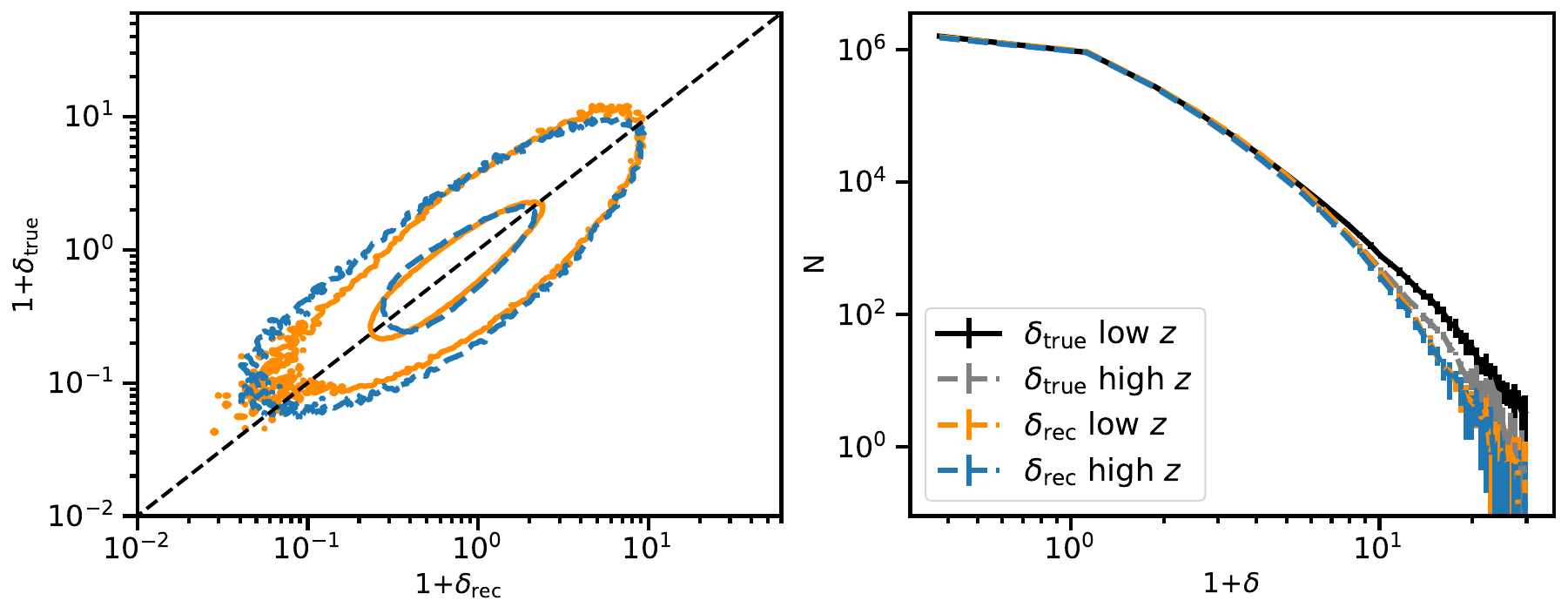}
    \caption{ The Same as Figure~\ref{fig:pix_com} but comparing the densities in the low-redshift (orange) and high-redshift (blue) region . }
    \label{fig:pix_com_diffz}
\end{figure*}
\begin{figure}
    \centering
    \includegraphics[trim=0cm 0cm 0cm 0cm, clip=True,width=1.0\columnwidth]{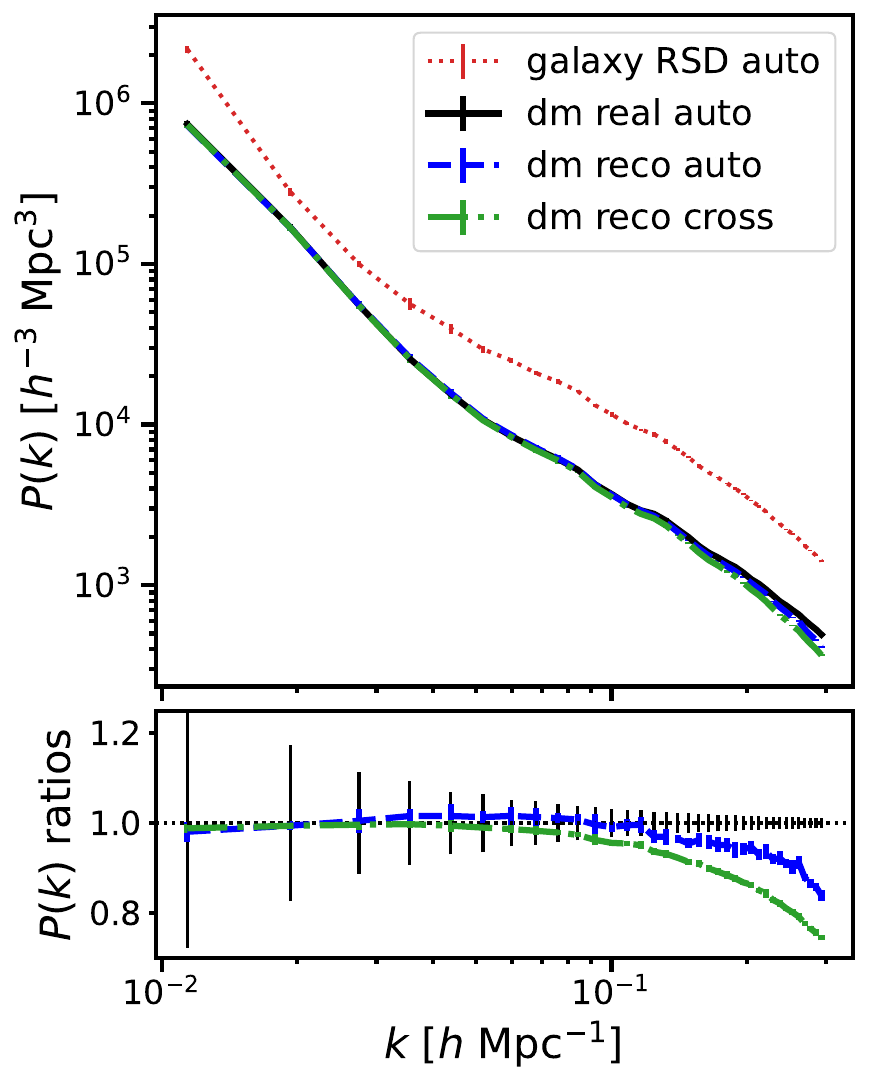}
    \caption{Comparison of $P(k)$ (upper panel) and $P(k)$ (lower panel) ratios. The black solid line and blue dashed line represent the autopower spectrum of the true and reconstructed density fields respectively,  while the green dotted-dashed line represents their cross-power spectra. For reference, the red dotted line shows the autopower spectra of the redshift-space galaxy field. The bottom panel shows the ratio between the reconstructed and true power spectra, with the error bars indicating the $\pm \sigma$ variance across the five test samples. The black solid error bars are computed using Eq.~(\ref{eq:pker}). }
    \label{fig:pk1d_den}
\end{figure}
\begin{figure}
    \centering
    \includegraphics[trim=0cm 0cm 0cm 0cm, clip=True,width=1.0\columnwidth]{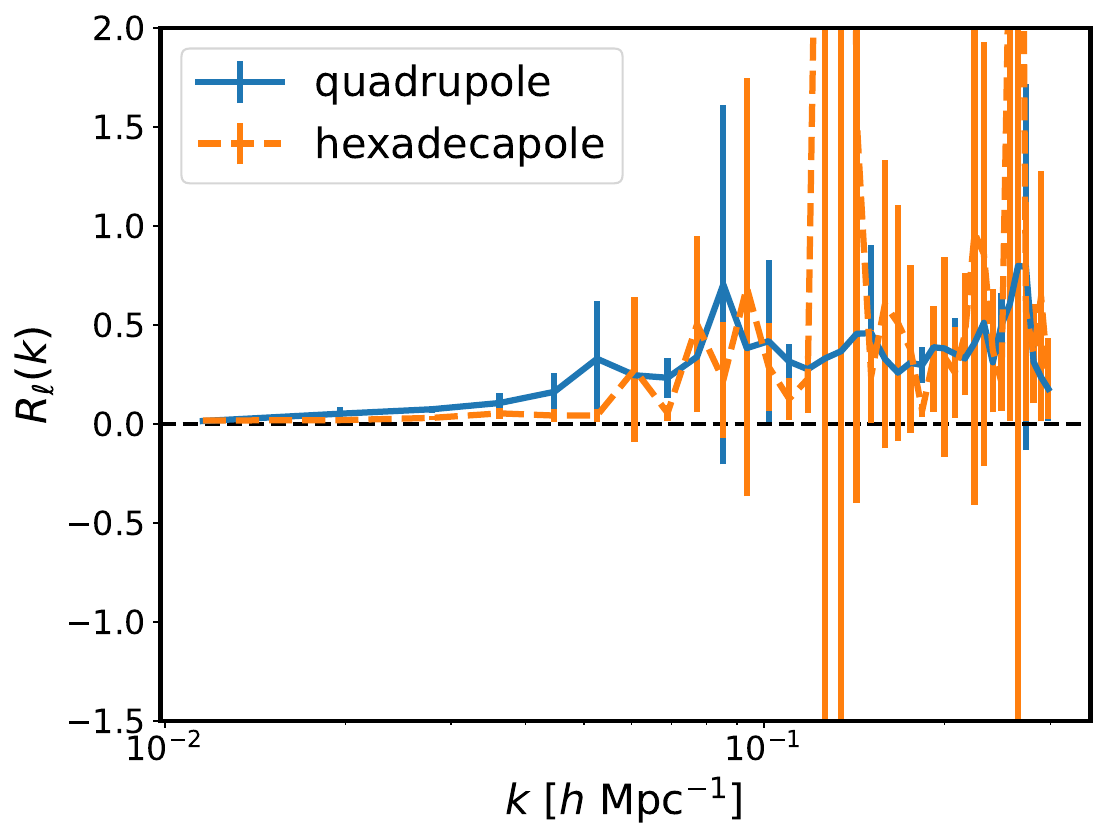}
    \caption{  The ratios $R_{\ell}(k)$ for the quadrupole (blue solid) and hexadecapole (orange dashed) power spectra. }
    \label{fig:pkl_den}
\end{figure}

\section{Results}\label{sec:res}
In this section, we present the reconstructed density and velocity fields derived from the mock DESI survey, by applying the resulting UNet model to five test samples. We also assess the reconstructed density field by validating its tidal field.

\subsection{Reconstructing the density field}

For visual comparison, Figure~\ref{fig:slice_den} shows the projected density field distribution in a slice of $1564 \times 1564 \times 9.16 \, h^{-1} \, \mathrm{Mpc}$. The Different panels correspond to the redshift-space galaxy distribution (upper left), the true dark matter density field $\delta_\mathrm{true}$ (upper right), the UNet-reconstructed density field $\delta_\mathrm{rec}$ (lower left), and the difference between $\delta_\mathrm{rec}$  and $\delta_\mathrm{true}$ (lower right). Compared with the true field, the reconstructed density field reveals recognizable and consistent LSSs, including clusters, filaments, and voids, indicating a successful reconstruction overall. Additionally, the results demonstrate resilience to boundary effects, as structures near the survey edges are effectively recovered. However, the difference map highlights discrepancies in high-density regions, particularly around clusters, where the reconstructed field deviates from the target. 

The left panel of Figure~\ref{fig:pix_com} illustrates the grid-to-grid density relationship between the reconstructed density field $1+\delta_\mathrm{rec}$ and the true field $1+\delta_\mathrm{true}$, represented by the blue solid line. For comparison, the result for the galaxy density field is also plotted. The three contours in each case represent thresholds encompassing 67\%, 95\%, and 99\% of the grid cells. The dashed lines indicate the ideal one-to-one correspondence. As expected, galaxies trace the underlying dark matter with a significant bias, leading to a relationship that deviates substantially from unity. However, after UNet reconstruction, the relationship becomes nearly linear and does not show significant bias. 

The right panel of Figure~\ref{fig:pix_com} shows the density distributions for galaxies (red dashed line), the reconstructed field (blue solid line), and the true field (black solid line). The error bars indicate the $\pm 1\sigma$ variance among the five test samples. After reconstruction, $\delta_\mathrm{rec}$ closely matches $\delta_\mathrm{true}$ in the regions where $1+\delta < 7$, but a systematic underestimation occurs for $1+\delta > 7$. This discrepancy likely arises from the imbalance in the number of grid cells between low- and high-density regions. As discussed in Section~\ref{sec:train}, the network initially prioritizes accurate predictions in low-density regions during early training stages, gradually improving high-density recovery at the cost of reduced correlation with the true field during later stages. The final model was selected by balancing scatter and correlation measures, which may contribute to the observed divergence. However, this issue affects only about 0.2\% of the grids within the $1+\delta > 7$ region. In comparison, the galaxy density distribution fails to trace the dark matter field accurately across both low- and high-density regimes.

Additionally,  Figure~\ref{fig:pix_com_diffz} presents a comparison between the low-redshift ($0.1 \lesssim z \lesssim 0.2$) and high-redshift ($0.3 \lesssim z \lesssim 0.4$) regions. In the left panel, the high-redshift region exhibits slightly larger scatter, highlighting the impact of lower galaxy density. In the right panel, the reconstructed density distribution underestimates high-density regions more significantly in the low-redshift region. This underestimation likely arises from higher galaxy clustering at low redshifts, which enhances nonlinear effects that the reconstruction method cannot fully correct, resulting in a suppression of high-density peaks. Overall, the method demonstrates robust performance, though further refinement is needed to better address nonlinear effects.

To evaluate the clustering, we compare the power spectra of the reconstructed density fields with the true fields. In this study, random samples were not used to correct for the survey's geometry or selection effects, as the focus is on evaluating the reconstruction's fidelity through direct comparison. This highlights the quality of the reconstruction but does not represent the true underlying clustering.  

The top panel of Figure~\ref{fig:pk1d_den} illustrates the power spectrum comparisons. The black solid line and blue dashed line represent the autopower spectra of the true and reconstructed density fields, respectively. For reference, the red dotted line shows the autopower spectrum of the redshift-space galaxy density field. The bottom panel shows the ratio between the reconstructed and true power spectra, with the error bars indicating the $\pm \sigma$ variance across the five test samples. By considering the cosmic variance due to a limited number of Fourier modes for a finite-volume survey, we present the theoretical error (black error bar) for the power spectrum\citep[e.g.][]{2021PhRvD.104d3528S} using
\begin{equation}\label{eq:pker}
    \Delta P(k) = \frac{P(k)}{\sqrt{N_k}}
\end{equation}
where $N_k$ denotes the number of independent $k$-modes available per bin. Compared to the halo $P(k)$, the reconstructed autocorrelation $P(k)$ values align with the true power spectrum within $1\sigma$ across the range of scales $k < 0.1 \, h \, \mathrm{Mpc}^{-1}$. The underestimation on scale $k > 0.1~h~\mathrm{Mpc}^{-1}$ is probably due to inaccurate reconstruction in the high-density region, as discussed above. The model's training objective, typically centered on minimizing global error metrics, may prioritize large-scale accuracy over small-scale fidelity, causing a modest degradation in the reconstructed small-scale power.
\begin{figure*}
    \centering
    \includegraphics[trim=0cm 0cm 0cm 0cm, clip=True,width=2.1\columnwidth]{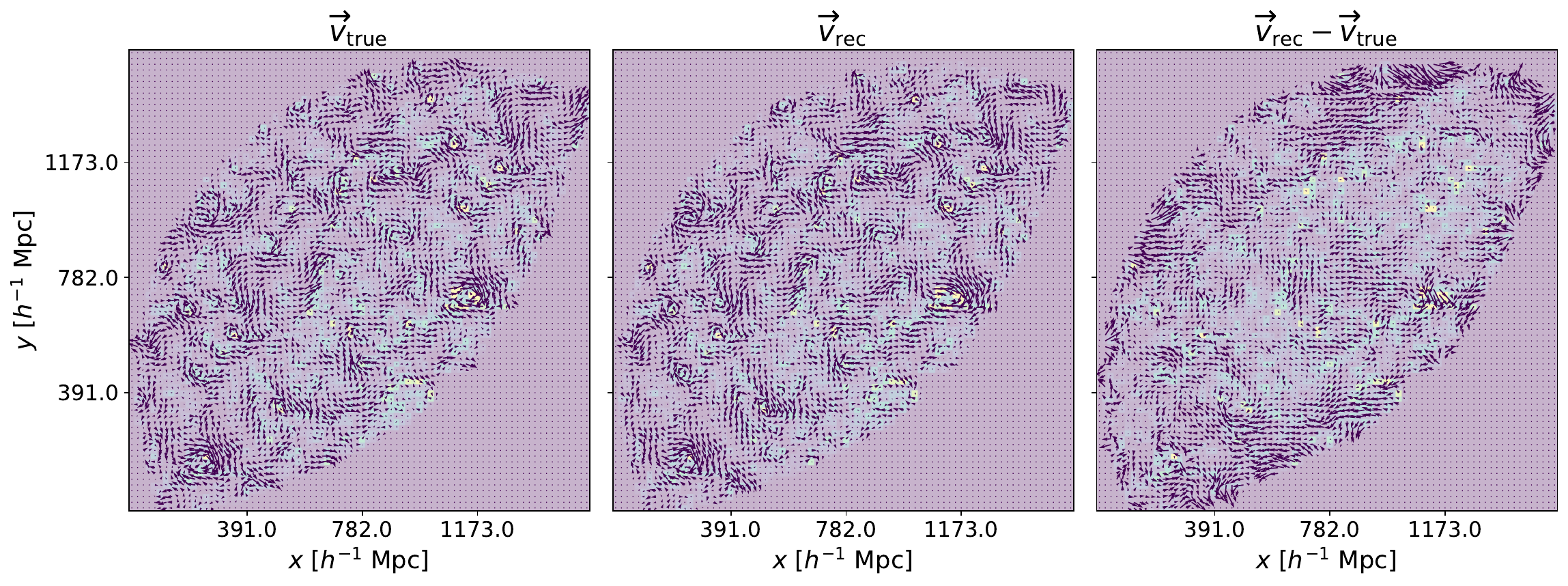}
    \caption{Comparison between the true and reconstructed dark matter velocity fields in a slice of dimensions $1564 \times 1564 \times 3.05 \, h^{-1} \, \mathrm{Mpc}$. The left panel shows the true velocity field $\vec{v}_\mathrm{true}$, directly extracted from the original simulation, while the middle panel presents the UNet-reconstructed field $\vec{v}_\mathrm{rec}$. The right panel shows the residuals, $\vec{v}_\mathrm{rec} - \vec{v}_\mathrm{true}$. In all panels, the arrows represent the velocity vectors, with their lengths proportional to the vector magnitudes. The underlying density field is also presented.}
    \label{fig:pix_vel}
\end{figure*}

To evaluate the concordance between the reconstructed and true density fields, we analyze the cross-correlation power spectrum. Unlike the autopower spectrum, which assesses the overall clustering amplitude, the cross-power spectrum directly measures how well the reconstructed field reproduces the phase and spatial distributions of the true field. High cross-correlation values indicate that the reconstruction effectively captures both large- and small-scale structures, making it a robust validation metric. In Figure~\ref{fig:pk1d_den}, the green dashed-dotted line represents the cross-power spectrum of the true and reconstructed fields, as well as the ratios of the reconstructed cross-power spectrum to the true one. The results show that the ratios align with unity within the $1\sigma$ range on scales $k < 0.1 \, h \, \mathrm{Mpc}^{-1}$, indicating a strong agreement between the two fields. At smaller scales ($k > 0.1 \, h \, \mathrm{Mpc}^{-1}$), the ratios fall below unity. Specifically, the reductions in the cross-correlation power spectrum are approximately 1.5\% at $k = 0.1 \, h \, \mathrm{Mpc}^{-1}$ and 20\% at $k = 0.3 \, h \, \mathrm{Mpc}^{-1}$. These reductions, though moderate, reflect challenges in accurately reconstructing high-density regions at smaller scales. Quantitatively, the average cross-correlation coefficient remains above 0.985 for $k < 0.1 \, h \, \mathrm{Mpc}^{-1}$, confirming the high fidelity of the reconstructed field for LSSs.

To evaluate the accuracy of the RSD corrections in the reconstructed density field, we compute the ratios of the multipole moments of the power spectra. These ratios compare the reconstructed and true power spectra, normalized by the difference between the redshift-space galaxy density power spectrum and the true power spectrum. The general formula for the ratio for any multipole $\ell$ is expressed as:
\begin{equation} 
R_{\ell}(k) = \frac{\lvert P_{\ell}^{\text{rec}}(k) - P_{\ell}^{\text{true}}(k) \rvert}{\lvert P_{\ell}^{\text{gal}}(k) - P_{\ell}^{\text{true}}(k) \rvert}, \end{equation}
where $P_{\ell}^{\text{rec}}(k)$ is the reconstructed dark matter power spectrum,
$P_{\ell}^{\text{true}}(k)$ is the true dark matter power spectrum, and
$P_{\ell}^{\text{gal}}(k)$ is the redshift-space galaxy density power spectra. The ratios are computed for both the quadrupole ($\ell=2$) and hexadecapole ($\ell=4$), providing a measure of the RSD correction performance at different scales.

Figure~\ref{fig:pkl_den} shows the ratios $R_{\ell}(k)$ for both the quadrupole ($\ell=2$) and hexadecapole ($\ell=4$) power spectra. At large scales, the ratios are close to zero, indicating that the RSD correction effectively restores the reconstructed power spectra to the true values. At small scales, the ratios are greater than zero but typically remain less than 0.5. This suggests that most of the RSDs have been successfully removed, although some residual distortions persist at smaller scales, probably due to less effective reconstruction in high-density regions. 

Based on the test results, we conclude that our network, despite using a small sample size, is capable of directly predicting the density field from the galaxy density field and effectively correcting for RSDs. This leads to a reliable reconstruction of the clustering mass density field. However, it is important to note that additional training data are still needed to enhance performance, particularly in high-density regions and on smaller scales.

\begin{figure*}
    \centering
    \includegraphics[trim=0cm 0cm 0cm 0cm, clip=True,width=2.1\columnwidth]{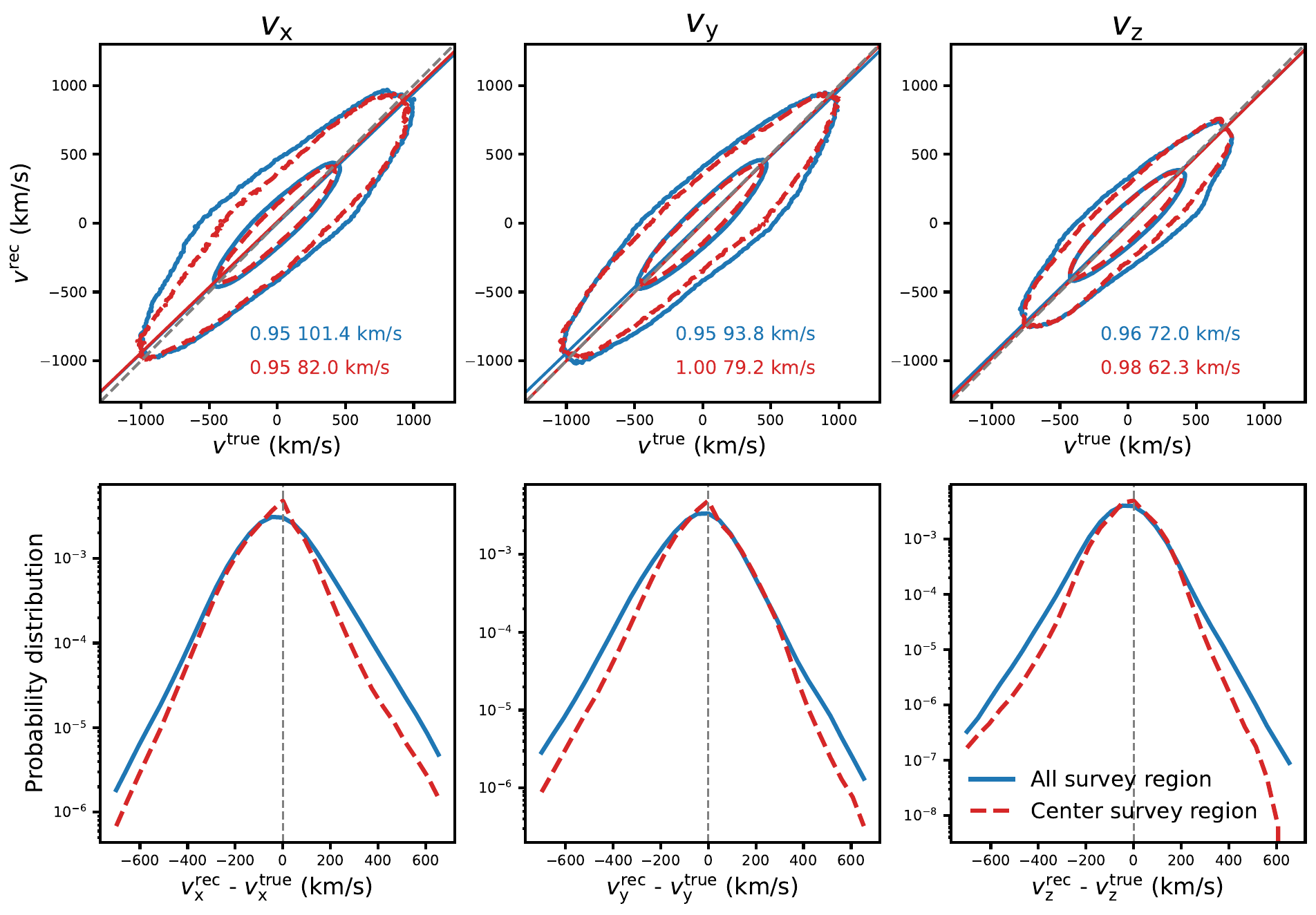}
    \caption{Upper panels: grid-to-grid velocity relations for the $x$-, $y$-, and $z$-axis velocity components, as shown in the different columns. The blue solid line corresponds to the results from the entire survey region, while the red dashed lines represent the results for the center survey region, defined as a sub-box within $391.0 , h^{-1} , \mathrm{Mpc} < x, y, z < 1173.0 , h^{-1} , \mathrm{Mpc}$. The contours represent 67\%, 95\%, and 99\% of the grid cells, with the best-fitting linear regressions shown as dashed lines.  The slope of the best-fitting relation and the scatter are also indicated in each panel. Lower panels: probability distributions of the velocity differences, $v_\mathrm{rec} - v_\mathrm{true}$.}
    \label{fig:hist_vel}
\end{figure*}
\begin{figure*}
    \centering
    \includegraphics[trim=0cm 0cm 0cm 0cm, clip=True,width=2.1\columnwidth]{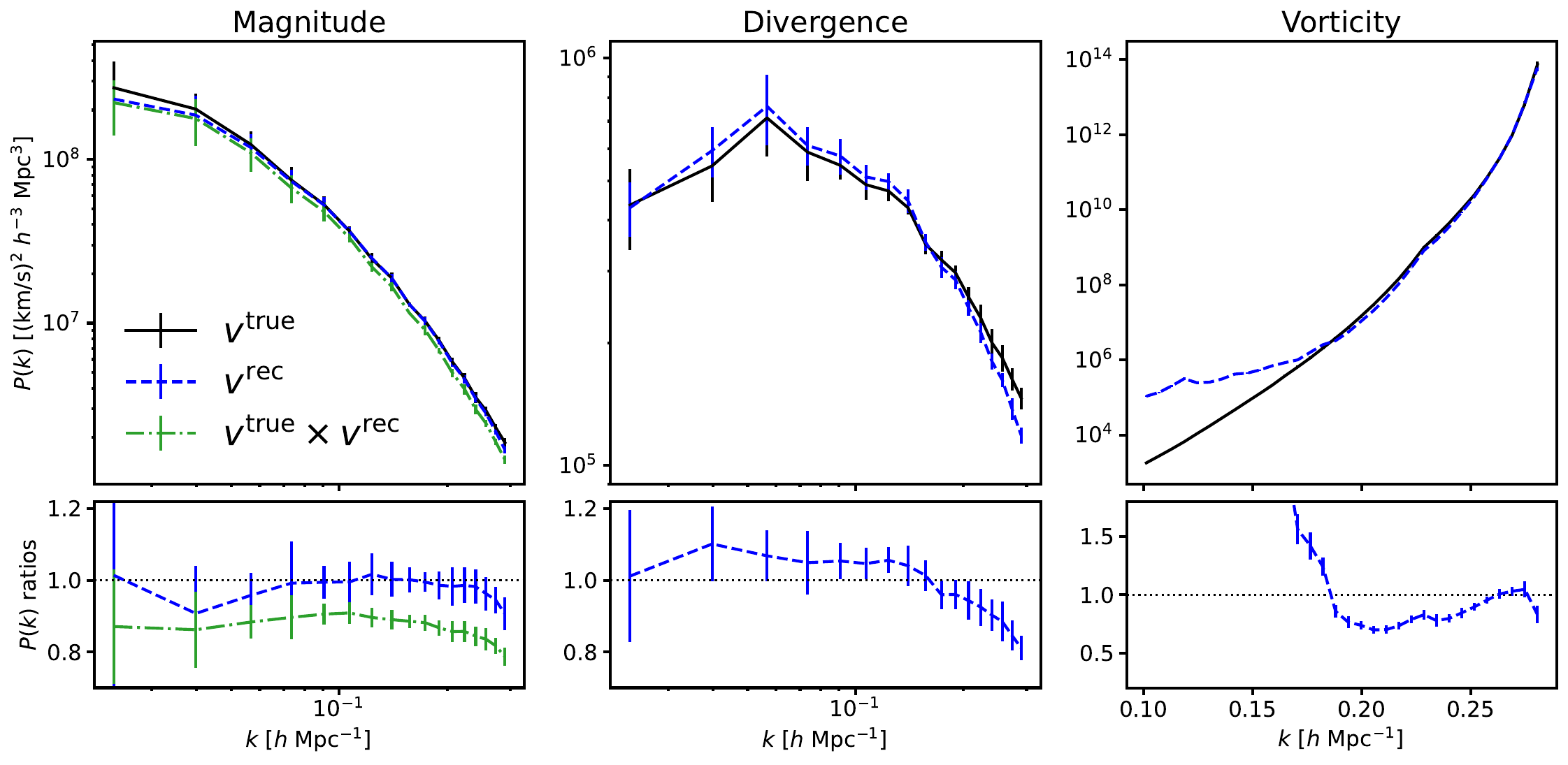}
    \caption{Comparison of the velocity field power spectrum. From left to right, the panels correspond to the velocity magnitude, divergence, and the vorticity, respectively. In each panel, the black solid and blue dashed lines represent the power spectra of the true and reconstructed velocity fields, respectively. The green dotted-dashed line shows the cross-correlation power spectra for the velocity magnitude between the true and reconstructed fields.}
    \label{fig:pk1d_vel}
\end{figure*}

\subsection{Reconstructing the velocity field}

Figure~\ref{fig:pix_vel} illustrates a comparison between the true and reconstructed dark matter velocity fields in a slice of dimensions $1564 \times 1564 \times 3.05 \, h^{-1} \, \mathrm{Mpc}$. The left panel shows the true velocity field, $\vec{v}_\mathrm{true}$, directly extracted from the original simulation, while the middle panel presents the UNet-reconstructed field, $\vec{v}_\mathrm{rec}$. The right panel shows the residuals, $\vec{v}_\mathrm{rec} - \vec{v}_\mathrm{true}$. In all panels, the arrows represent the velocity vectors, with their lengths proportional to the vector magnitudes. The underlying density field is also presented. The reconstructed field $\vec{v}_\mathrm{rec}$ closely matches $\vec{v}_\mathrm{true}$, successfully capturing both large-scale coherent flows and small-scale turbulent features. This demonstrates UNet's capability to recover the true velocity structure across a wide range of scales. However, the residual map of the right panel highlights discrepancies mainly near the survey boundaries. These deviations are likely caused by boundary effects, as the reconstruction model struggles to fully account for the lack of surrounding information at the edges of the domain. Despite these boundary-related residuals, the overall performance of the reconstruction remains robust, with the primary flow patterns and magnitudes being well reproduced.

The upper panels of Figure~\ref{fig:hist_vel} show the grid-to-grid velocity relations for the $x$-, $y$-, and $z$-axis velocity components, as shown in different columns. The blue solid line corresponds to the results from the entire survey region, while the red dashed lines represent the results for the center survey region, defined as a sub-box within $391.0 , h^{-1} , \mathrm{Mpc} < x, y, z < 1173.0 , h^{-1} , \mathrm{Mpc}$. The contours represent 67\%, 95\%, and 99\% of the grid cells, with the best-fitting linear regressions shown as dashed lines. The slopes of the best-fit relations are close to unity for both regions, with values around $0.95$–$1.00$, indicating strong agreement between the reconstructed $v_\mathrm{rec}$ and true velocities $v_\mathrm{true}$. The scatter values, reflecting the variances from the best-fit lines, are smaller in the center region, demonstrating improved reconstruction precision due to reduced boundary effects. The lower panels of Figure~\ref{fig:hist_vel} illustrate the probability distributions of the velocity differences, $v_\mathrm{rec} - v_\mathrm{true}$. These distributions peak nicely at zero, and the center region shows narrower peaks, further verifying its superior accuracy. In contrast, the broader distributions over the entire survey region underscore the challenges posed by boundary effects and edge distortions. Overall, the results demonstrate the reliability of the reconstruction, with notable improvements in the center survey region compared to the full survey.

To further evaluate the reconstructed velocity field, we calculate the power spectrum, as shown in Figure~\ref{fig:pk1d_vel}. In each panel, the black solid line represents the power spectrum of the true velocity field, while the blue dashed line corresponds to the reconstructed field. Additionally, the green dotted-dashed line shows the cross-correlation power spectrum for the velocity magnitude between the true and reconstructed fields, which serves as a measure of their concordance. Overall, the reconstructed velocity field demonstrates strong agreement with the true field across the three components, with some deviations observed.

The cross-power spectrum for the velocity magnitude is systematically lower than unity by approximately 10\% on all scales. This discrepancy can be attributed to two factors: (1) survey boundary effects, which truncate the velocity field at the edges of the survey volume and introduce artifacts, particularly on large scales, weakening the correlation between the true and reconstructed modes; and (2) training limitations, as the reconstruction model, typically trained to minimize global error metrics, prioritizes large-scale accuracy at the cost of small-scale fidelity, resulting in suppressed power on smaller scales. For the vorticity power spectrum, the discrepancy at scales $k < 0.2 \, h \, \mathrm{Mpc}^{-1}$ may arise from the intrinsic nature of the vorticity signal. As vorticity is primarily sourced from small-scale, nonlinear velocity interactions, the reconstruction method may face challenges in accurately reproducing these subtle features at relatively large scale where the signal-to-noise ratio is inherently low. However, at smaller scales $k > 0.2 \, h \, \mathrm{Mpc}^{-1}$, the reconstructed vorticity shows better agreement with the true field, reflecting the model's ability to capture localized turbulence. These results underscore the robust overall performance of the reconstruction method, while also highlighting the limitations introduced by survey boundaries and algorithmic challenges.
\begin{figure*}
    \centering
    \includegraphics[trim=0cm 0cm 0cm 0cm, clip=True,width=2.1\columnwidth]{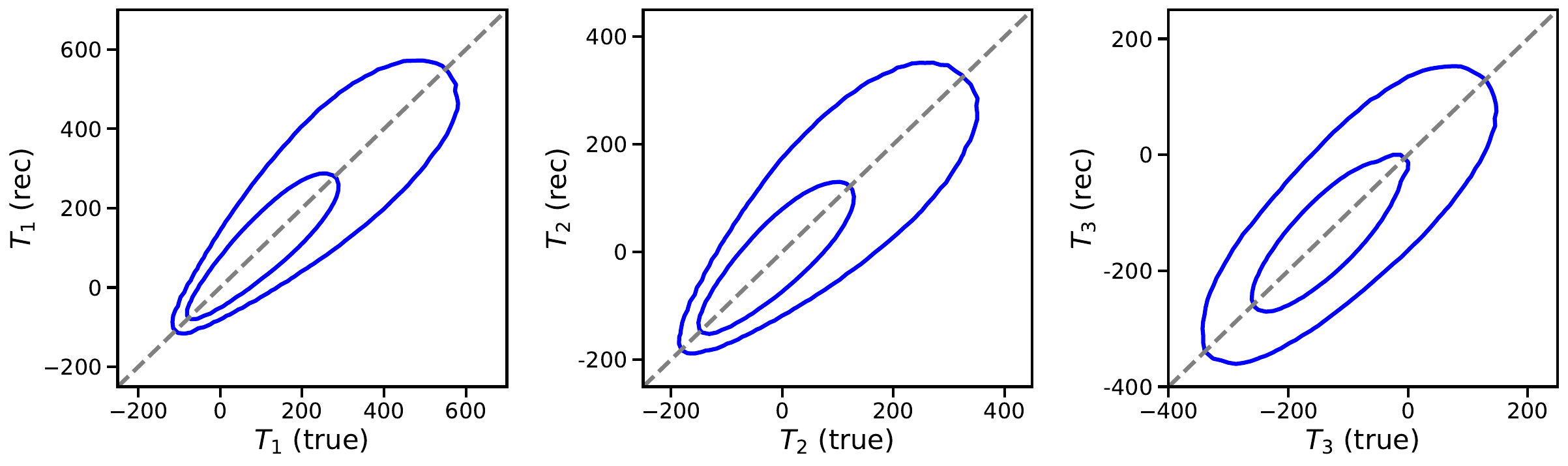}
    \caption{Eigenvalues of the reconstructed tidal tensor versus those obtained from the true mass density field. The contours for each result encompass 95\% and 99\% of the grid cells.}
    \label{fig:pix_tidal}
\end{figure*}
\begin{figure*}
    \centering
    \includegraphics[trim=0cm 0cm 0cm 0cm, clip=True,width=2.1\columnwidth]{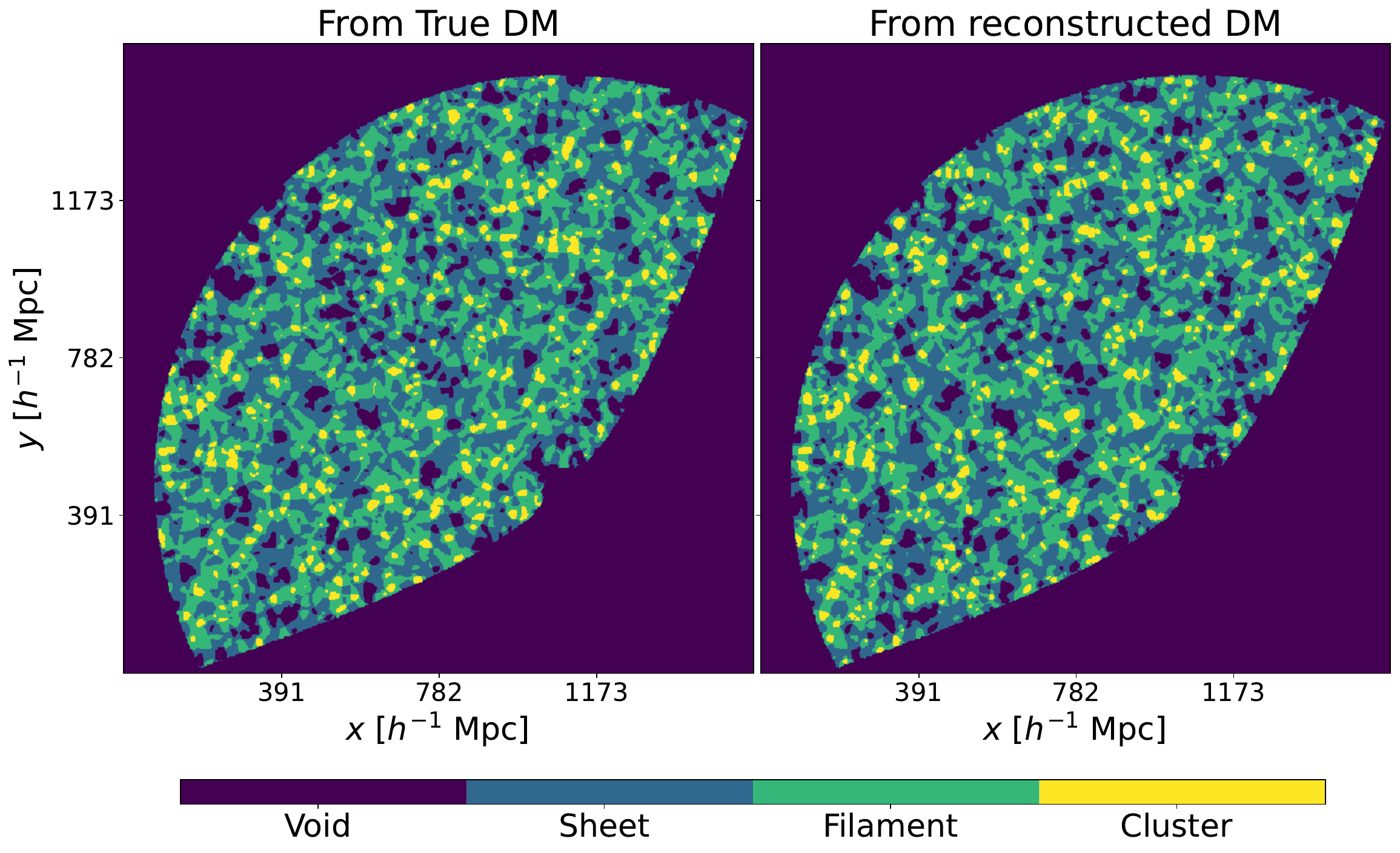}
    \caption{Comparison of the classification of the LSS between the true mass density field (left panel) and the UNet-reconstructed density field (right panel) in a slice of $1564 \times 1564 \times 3.05 \mpch$. The yellow, yellow-green, green, and black regions show grid cells located at structures classified as clusters, filaments, sheets, and voids, respectively.}
    \label{fig:slices_tidal}
\end{figure*}
\begin{figure}
    \centering
    \includegraphics[trim=0cm 0cm 0cm 0cm, clip=True,width=1.0\columnwidth]{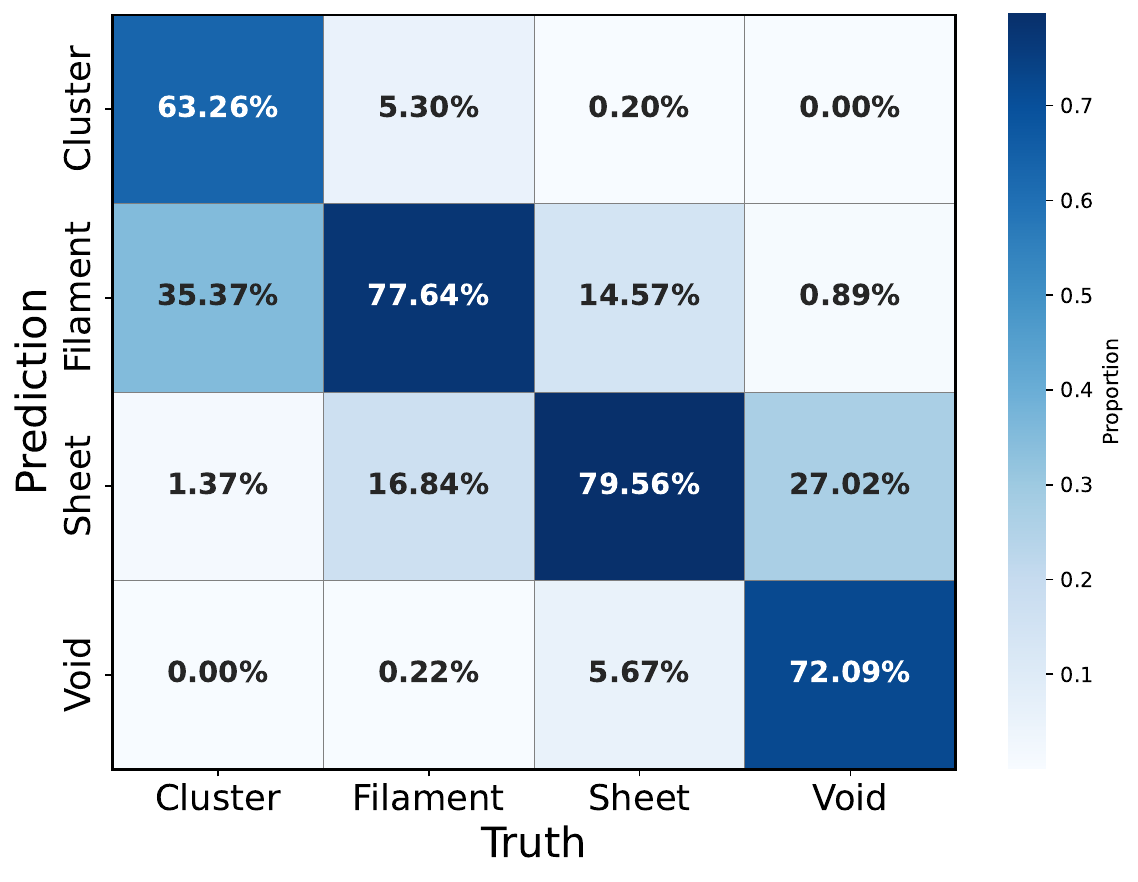}
    \caption{Normalized confusion matrix. Each row corresponds to a predicted class derived from the UNet-reconstructed dark matter density field, while each column corresponds to the true class based on the original simulated density field. The matrix entries represent classification probabilities, obtained by normalizing the values within each column such that the total for each true class sums to 1.}
    \label{fig:confmatrix}
\end{figure}

\subsection{Reconstructing the tidal field}\label{sec:tidal}

We now proceed with the reconstruction of the large-scale tidal field using the tidal tensor, $T_{ij}$, defined as
\begin{equation}
    T_{ij} = \frac{\partial^2\phi}{\partial x_i \, \partial x_j},
\end{equation}
where $i$ and $j$ are indices taking values 1, 2, or 3, and $\phi$ is the peculiar gravitational potential. We compute $\phi$ from the mass density field using the Poisson equation
\begin{equation}
    \nabla^2 \phi = 4\pi G \bar{\rho} \delta,
\end{equation}
where $\bar{\rho}$ is the average density of the Universe. The tidal tensor is then diagonalized to obtain the eigenvalues $\lambda_1 \geqslant \lambda_2 \geqslant \lambda_3$ at each grid point.

Figure~\ref{fig:pix_tidal} compares the eigenvalues of the tidal field tensor ($T_1$, $T_2$, $T_3$) derived from the UNet-reconstructed mass density field to those obtained from the true density field of the mock DESI survey. The relation plots for each eigenvalue reveal a strong alignment along the diagonal, demonstrating excellent agreement between the reconstructed and true tidal fields. The contours, representing 95\% and 99\% of the grid cells, further confirm that the reconstructed eigenvalues closely match the true values, validating the accuracy of the reconstruction method. This result is consistent with our previous work (Wang24), which showed that the UNet-based reconstruction method significantly outperforms traditional halo-based approaches \citep{2009MNRAS.394..398W} in recovering unbiased tidal fields.

The eigenvalues of the tidal field can be used to classify the LSS into four categories based on their local morphology:
\begin{itemize} \item Cluster: grid cells with three positive eigenvalues;
\item Filament: grid cells with one negative and two positive eigenvalues;
\item Sheet: grid cells with two negative and one positive eigenvalues;
\item Void: grid cells with three negative eigenvalues.
\end{itemize}
Figure~\ref{fig:slices_tidal} shows the LSS classifications derived from the true mass density field (left panel) and the UNet-reconstructed density field (right panel). The visual comparison highlights a remarkable consistency between the two panels, with the spatial distributions of structures, including clusters, filaments, sheets and voids, closely matching across the slice. This indicates that the UNet-reconstructed density field effectively preserves the topological connectivity and morphological details of the cosmic web. 

Figure~\ref{fig:confmatrix} presents the normalized confusion matrix evaluating the classification performance of the UNet-reconstructed LSS field. The matrix compares the predicted LSS types derived from the reconstructed dark matter density field against the ground-truth classification from the original simulation. Each row corresponds to the predicted class, while each column corresponds to the true class, and the matrix entries represent the column-normalized classification probabilities. The diagonal elements indicate the proportion of correctly predicted classes, highlighting that sheets and filaments are reconstructed with relatively high fidelity (79.56\% and 77.64\%, respectively), followed by voids (72.09\%) and clusters (63.26\%). Clusters are frequently misclassified as filaments (35.37\%), reflecting the network’s tendency to underestimate high-density peaks due to their sparsity and sharp gradients. Voids, while generally well recovered, show moderate confusion with sheets (27.02\%), likely due to boundary smoothing in the reconstructed field, which reduces the contrast between underdense void interiors and adjacent sheet-like structures. Overall, the confusion matrix confirms that the UNet model robustly reconstructs the cosmic web morphology, especially in filamentary and sheet-like environments, while revealing relatively higher level of challenge in terms of reconstructing dense cluster regions.

% These results demonstrate that the UNet model effectively captures the spatial and topological features of the cosmic web, particularly for well-defined environments such as voids and dense sheets, while transitional zones between filaments and clusters remain more challenging due to their intermediate densities and complex morphology.}
% These trends are consistent with the reconstruction bias observed in Figure 7, where intermediate-density regions are recovered more accurately than the extremes.

% However, small discrepancies are observed, including a slight underestimation of the void volume near the boundaries and misrepresentation of filamentary or sheet regions. These differences likely arise from boundary effects in the survey region, which can truncate the density field near the edges, as well as the inherent challenges in predicting intermediate-density environments, where the contrast between sheets and filaments is less distinct. Despite these minor biases, the UNet reconstruction provides a robust and reliable representation of the tidal field, enabling accurate LSS classification for cosmological studies.

\section{Discussion}
The reconstruction framework presented in this work demonstrates robust performance in recovering the overall dark matter density and velocity fields. However, 
it exhibits scale- and environment-dependent limitations that  that may affect its applicability to certain scientific analyses.

One such limitation arises in the reconstruction of high-density peaks, where the model systematically underestimates the true density in regions with $1 + \delta > 7$. Although such regions constitute only about 0.2\% of the total grid volume, they correspond to massive cluster-scale structures that host significant cosmological information. This underestimation likely stems from the rarity of these regions in the training data and the intrinsic limitations of convolutional networks in capturing sharp density contrasts. As a result, analyses that require accurate reconstructions of cluster morphology or mass, such as environmental dependence studies (see Section~\ref{sec:tidal}), non-Gaussianity analyses \citep[e.g.,][]{2013JCAP...08..004S}, gravitational lensing \citep[e.g.,][]{2013SSRv..177...75H}, and Sunyaev–Zel’dovich cluster catalogs \citep[e.g.,][]{2021ApJS..253....3H}, can be adversely affected.

In addition, the power spectrum analysis reveals a scale-dependent discrepancy at $k>0.1~h~\mathrm{Mpc}^{-1}$, where the reconstructed field underestimates the true power by up to ahout 20\%. This deviation, consistent with the suppressed densities in high-density regions, indicates a limitation in recovering nonlinear clustering information. While the reconstruction remains reliable for LSS statistics, it biases the small-scale power spectrum, limiting its utility in testing nonlinear gravitational models \citep[e.g.,][]{2019MNRAS.488.2121C} and quantifying baryonic feedback effects\citep[e.g.,][]{2024MNRAS.533..621P}. Inaccuracies in the power spectrum can also compromise the accuracy of cosmological parameter inference based on small-scale power\citep[e.g.,][]{2021JCAP...11..038O}.  Additional consequences may include reduced fidelity in reconstructing initial conditions or in generating forward-modeled observables \cite[e.g.,][]{2016ApJ...831..164W}.  

The reconstructed velocity field also exhibits a systematic suppression in amplitude across all scales, with a roughly 10\% deficit in the cross-power spectrum relative to the true velocity field. This underestimation can impact a range of scientific applications that rely on accurate 3D velocity information. For example, kinetic Sunyaev–Zel’dovich tomography, which cross-correlates reconstructed velocity fields with cosmic
microwave background temperature maps\citep{2011MNRAS.413..628S}, requires the faithful recovery of both amplitude and direction to avoid biasing the inferred baryon distribution. Similarly, the use of velocity fields to trace bulk flows \citep{2020MNRAS.498.2703B}, analyze void dynamics\citep{2016MNRAS.461.4013C}, or reconstruct initial conditions\citep{2013MNRAS.430..888D} depends on preserving coherent structures across a wide range of scales. The suppressed velocity power may therefore reduce the effectiveness of these analyses. Improving velocity reconstruction may require velocity-specific loss functions and enhanced boundary treatment to support precision applications in velocity-based cosmology.

Therefore, the current reconstruction framework performs well in capturing large-scale and intermediate-density structures. However, its accuracy degrades in regimes characterized by nonlinear structures and velocity complexity. Extending its applicability to these domains will likely require improvements in the training methodology and model architecture, including adaptive loss functions, focused data augmentation, and multiscale designs to better capture nonlinear structure and velocity information. We will pursue these enhancements in future work to broaden the framework’s scientific utility.

% Similarly, momentum field reconstruction plays a key role in cross-correlation studies with lensing or intensity mapping surveys.

\section{Summary and Conclusion}\label{sec:sum}

In this study, we have reconstructed the MTV fields using a DESI-like BGS based on a UNet convolutional neural network. Our approach transforms galaxy density fields into the underlying dark matter properties by exploiting UNet's capability to establish precise field-to-field mappings. This method significantly enhances the recovery of LSSs and captures the intricate morphology of cosmic web features.

We first generated a DESI-like survey using the Jiutian N-body simulation, which provides high resolution in accordance with Planck2018 cosmological parameters. From the survey, we selected BGS galaxies and underlying dark matter particles within a redshift range of $0.1 < z < 0.4$. The resulting light cone was carefully trimmed to DESI's sky coverage of approximately $9625.23 \, \mathrm{deg^2}$, providing a realistic and comprehensive dataset for both galaxy and dark matter distributions to train and test our models.

We successfully reconstructed the mass density field, achieving close agreement with the true field across LSSs such as clusters, filaments, and voids. The cross-correlation power spectrum demonstrates a reconstruction accuracy exceeding 98.5\% for scales with $k < 0.1 \, h \, \mathrm{Mpc}^{-1}$. Moderate discrepancies, with reductions of 1.5\% at $k = 0.1 \, h \, \mathrm{Mpc}^{-1}$ and 20\% at $k = 0.3 \, h \, \mathrm{Mpc}^{-1}$, were observed in high-density regions due to challenges in reconstructing small-scale features. Furthermore, we evaluated the performance of the RSD correction using multipole moment ratios. These results confirm that the reconstructed field effectively restores the clustering signal by removing most of the RSD effects, achieving strong consistency with the true field, particularly at large scales.

Our velocity field reconstruction accurately recovered large-scale coherent flows and small-scale turbulent features. Grid-to-grid comparisons yielded slopes close to unity (0.95–1.00) with reduced scatter in central regions, highlighting robust performance. The cross-correlation power spectrum was systematically underestimated by approximately 10\% across all scales, primarily due to boundary effects and small-scale reconstruction challenges. Despite these limitations, the results emphasize the model's ability to retain the structure of the velocity field on a wide range of scales.

We reconstructed the tidal field by computing the eigenvalues of the tidal tensor and found strong agreement between the reconstructed and true fields. Our approach accurately preserved LSS classifications, including clusters, filaments, sheets, and voids, while maintaining the morphological connectivity of the cosmic web. Minor biases were observed, such as a slight underestimate of the void volume near the boundaries and a misrepresentation of filamentary or sheet regions. These discrepancies are mainly attributed to boundary effects and the challenges of predicting intermediate-density environments. In general, the results confirm the ability of the method to reliably recover the structure of the tidal field.

We have shown that a UNet-based deep learning framework can effectively reconstruct MTV fields from galaxy samples. While the method performs well overall, its accuracy declines in high-density regions and shows moderate suppression in velocity field reconstruction. These limitations point to directions for future improvement. Nonetheless, the approach provides a promising tool for LSS analysis in forthcoming observational surveys.

% We have demonstrated that deep learning, specifically a UNet-based architecture, is an effective tool for reconstructing cosmological fields from galaxy samples. Our method consistently achieved high fidelity across density, velocity, and tidal fields, providing a reliable framework for LSS analysis. This study highlights potential applications in observational surveys and enhances our understanding of dark matter distributions and dynamics.

%% Also note that the akcnowlodgment environment does not support long amounts of text. If you have a lot of people and institutions to acknowledge, do not use this command. Instead, create a new \section{Acknowledgments}.
% \begin{acknowledgments}

\section{Data AVAILABILITY}
The dataset of the 3D fields of the simulated DESI BGS galaxy and dark matter distributions will be made available through the PaperData Repository of the National Astronomical Data Center of China: \url{https://doi.org/10.12149/101654}.

\section{Acknowledgments}
This work is supported by the National SKA Program of China (2022SKA0110200 and 2022SKA0110202), the National Key R\&D Program of China (2023YFA1607800, 2023YFA1607801, 2023YFA1607804), the National Natural Science Foundation of China (Nos.12103037, 12261141691, 12373005), 111 project No. B20019, Shanghai Natural Science Foundation (Nos. 19ZR1466800, 24DX1400100, 23JC1410200) and Zhangjiang special support (No. ZJ2023-ZD-003). We acknowledge the science research grants from the China Manned Space Project (Grant No. CMS-CSST-2021-A02, CMS-CSST-2021-A03, CMS-CSST-2021-B01) and the Fundamental Research Funds for the Central Universities (Grant No. ZYTS25248, XJS221312). This work is supported by the State Key Laboratory of Dark Matter Physics and the Young Data Scientist Program of the China National Astronomical Data Center (No.NADC2025YDS-01). H.W. acknowledges support from the National Natural Science Foundation of China (NSFC, Nos. 12192224) and CAS Project for Young Scientists in Basic Research, Grant No. YSBR-062. This work is supported by Natural Science Basic Research Program of Shaanxi (Program No. 2025JC-YBMS-016). The authors acknowledge the Beijing Super Cloud Center (BSCC) for providing HPC resources that have contributed to the research results reported in this paper. This work is also supported by the High-Performance Computing Platform of Xidian University. 
% \end{acknowledgments}

%% To help institutions obtain information on the effectiveness of their 
%% telescopes the AAS Journals has created a group of keywords for telescope 
%% facilities.
%
%% Following the acknowledgments section, use the following syntax and the
%% \facility{} or \facilities{} macros to list the keywords of facilities used 
%% in the research for the paper.  Each keyword is check against the master 
%% list during copy editing.  Individual instruments can be provided in 
%% parentheses, after the keyword, but they are not verified.

\vspace{5mm}
% \facilities{HST(STIS), Swift(XRT and UVOT), AAVSO, CTIO:1.3m,
% CTIO:1.5m,CXO}

%% Similar to \facility{}, there is the optional \software command to allow 
%% authors a place to specify which programs were used during the creation of 
%% the manuscript. Authors should list each code and include either a
%% citation or url to the code inside ()s when available.

\software{\texttt{Astropy} \citep{astropy_collaboration_astropy_2018, astropy_collaboration_astropy_2022}, \texttt{NumPy} \citep{van_der_walt_numpy_2011, harris_array_2020}, \texttt{Matplotlib} \citep{hunter_matplotlib_2007}, \texttt{Tensorflow}\citep{tensorflow2015-whitepaper}, \texttt{Pylians}\citep{Pylians}}

\appendix
% \section{Comparison of the clustering between mock sample and DESI}

\section{Testing on FA effect}\label{sec:FA}
\begin{figure}
    \centering
    \includegraphics[trim=0cm 0cm 0cm 0cm, clip=True,width=0.9\columnwidth]{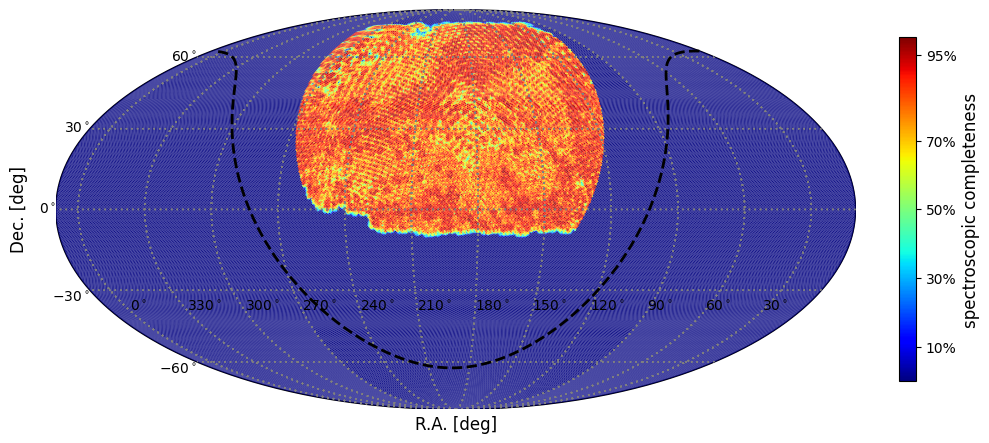}
    \caption{Completeness map of the DESI survey footprint simulated using the four-pass observation strategy.}
    \label{fig:FA}
\end{figure}

To assess the potential impact of FA on the reconstruction, we simulate the FA effect using a four-pass observation strategy, consistent with the DESI survey design \citep{2023AJ....165..253H, 2025JCAP...04..074B}. Each pass corresponds to a separate tiling of the survey footprint, targeting available galaxies while accounting for fiber collisions and survey geometry. After applying this simulation, the average completeness across the survey area is approximately 80\%, reflecting the partial sampling introduced by the FA process. Figure~\ref{fig:FA} shows the resulting completeness map for the DESI footprint, highlighting variations in coverage due to the tiling strategy and the FA algorithm. This map illustrates the nonuniform spatial sampling that can introduce systematic biases in clustering measurements and reconstruction performance.

\begin{figure*}
    \centering
    \includegraphics[trim=0cm 0cm 0cm 0cm, clip=True,width=0.9\columnwidth]{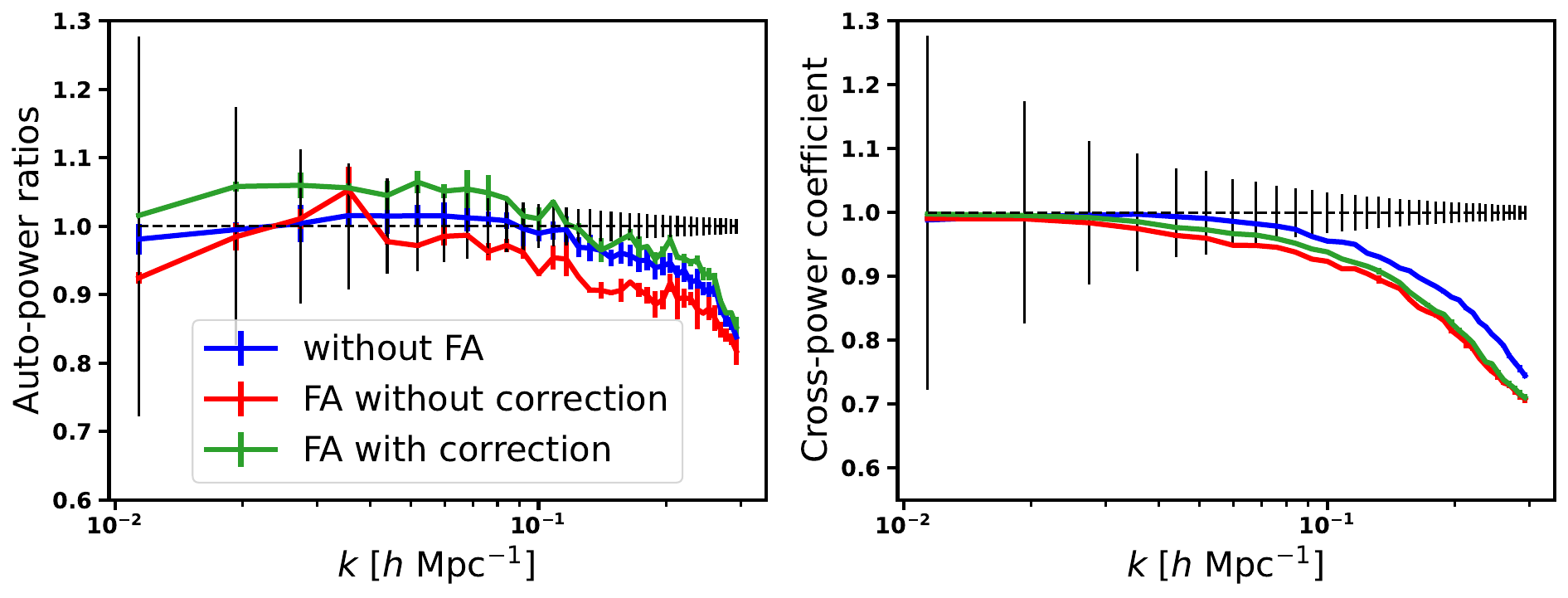}
    \caption{Ratios of the auto- (left panel) and cross-power (right panel) spectra between the reconstructed and true density fields, illustrating the impact of the FA effect. The blue lines show results without the FA effect and the red lines include the FA effect. The green lines represent a simple correction using nearest-neighbor galaxies.}
    \label{fig:Pk_FA}
\end{figure*}

We then retrain the model using the sample with FA included. Figure~\ref{fig:Pk_FA} shows the ratios of the auto- and cross-correlation power spectra. The blue line represents results without the FA effect, while the red lines show results with the FA effect included. The green lines illustrate a simple correction using the nearest-neighbor galaxies. At large scales ($k \lesssim 0.1\,h\,\mathrm{Mpc}^{-1}$), all curves remain close to unity, indicating accurate reconstruction. However, at smaller scales, the FA effect introduces a noticeable suppression in the autopower spectrum (red lines), and the cross-power spectrum ratio also declines. The nearest-neighbor correction partially mitigates this suppression, as seen in the green lines. These results demonstrate that while the FA effect does not significantly impact LSSs, it biases small-scale clustering and highlights the need for more sophisticated corrections in future work. Nevertheless, we find that at small scales, the observed underestimation is dominated more by the limitations of the UNet prediction than by the FA effect itself.


\section{Comparison between Separately Trained and Unified Models}\label{sec:compare_mods}

To assess the potential benefits of unified training, we conduct an additional experiment in which a single network is designed to reconstruct both the density and velocity fields simultaneously. This configuration contrasts with our original approach, where separate models were trained for each field. Due to hardware memory limitations, we performed this comparison using data cubes of size $256^3$ rather than the original $512^3$ cubes employed in our main analysis. 

\begin{figure*}
    \centering
    \includegraphics[trim=0cm 0cm 0cm 0cm, clip=True,width=1.0\columnwidth]{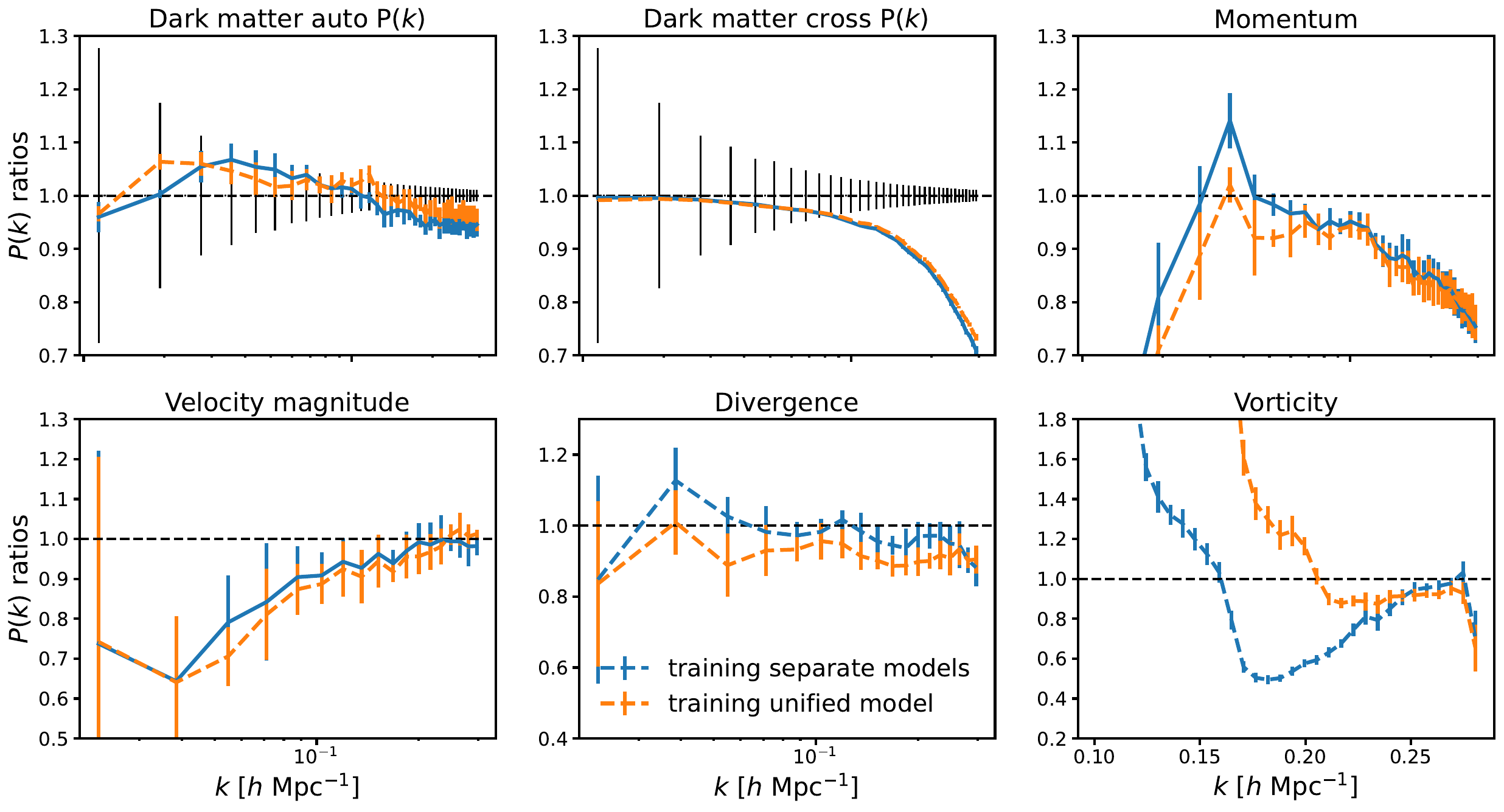}
    \caption{Comparison of the reconstruction performance between the separately trained models (blue lines) and the unified model (orange lines). As indicated in the panel titles, each subplot corresponds to one of the following diagnostic metrics: (1) the autopower spectrum of the dark matter field; (2) the cross-power spectrum between the reconstructed and true dark matter fields; (3) the momentum field power spectrum; (4) the velocity magnitude power spectrum, (5) the velocity divergence power spectrum; and (6) the vorticity power spectrum.}
    \label{fig:pk_compare_mods}
\end{figure*}

In Figure~\ref{fig:pk_compare_mods}, we present a comparative analysis of the reconstruction performance between the separately trained models (blue lines) and the unified model (orange lines). Our results indicate that the overall reconstruction performance is largely consistent between the separate and unified training strategies, although noticeable discrepancies appear in certain metrics, such as the divergence and vorticity power spectra. In particular, the separately trained models demonstrate improved accuracy in reconstructing the divergence field, exhibiting lower scatter and reduced power suppression. Notably, these models still recover coherent structures across both the density and velocity fields, suggesting that feature consistency is well maintained even without joint training. This implies that despite the absence of shared parameters, separate training does not compromise the alignment between the reconstructed fields under the current experimental setup. 

While it is generally expected that a unified model would produce more consistent features between the reconstructed density and velocity fields, the results indicate that under current memory constraints, training separate models remains the preferred strategy for maximizing accuracy across a wide range of scales. Future implementations on GPUs with larger memory capacity may revisit the unified approach as a viable and potentially more efficient alternative.

%   

%% Appendix material should be preceded with a single \appendix command.
%% There should be a \section command for each appendix. Mark appendix
%% subsections with the same markup you use in the main body of the paper.

%% Each Appendix (indicated with \section) will be lettered A, B, C, etc.
%% The equation counter will reset when it encounters the \appendix
%% command and will number appendix equations (A1), (A2), etc. The
%% Figure and Table counter will not reset.

% \appendix

% \section{Appendix information}

% Appendices can be broken into separate sections just like in the main text.

%% For this sample we use BibTeX plus aasjournals.bst to generate the
%% the bibliography. The sample631.bib file was populated from ADS. To
%% get the citations to show in the compiled file do the following:
%%
%% pdflatex sample631.tex
%% bibtext sample631
%% pdflatex sample631.tex
%% pdflatex sample631.tex

\bibliography{sample631}{}
\bibliographystyle{aasjournal}

%% This command is needed to show the entire author+affiliation list when
%% the collaboration and author truncation commands are used.  It has to
%% go at the end of the manuscript.
%\allauthors

%% Include this line if you are using the \added, \replaced, \deleted
%% commands to see a summary list of all changes at the end of the article.
%\listofchanges

\end{document}